\documentclass{aastex}
\usepackage{emulateapj5}
\usepackage{onecolfloat5}
\usepackage{apjfonts}
\usepackage{epsfig} 

\def\myputfigure#1#2#3#4#5%
{\vskip#5pt\makebox[0pt]{\hskip#2in
\includegraphics[width=#3\textwidth]{#1}}\vskip#4pt\hfill}

\def\gsim{\;\rlap{\lower 2.5pt
 \hbox{$\sim$}}\raise 1.5pt\hbox{$>$}\;}
\def\lsim{\;\rlap{\lower 2.5pt
   \hbox{$\sim$}}\raise 1.5pt\hbox{$<$}\;}
\newcommand{\be}{\begin{equation}}
\newcommand{\beq}{\begin{equation}}
\newcommand{\ba}{\begin{eqnarray}}
\newcommand{\ee}{\end{equation}}
\newcommand{\eeq}{\end{equation}}
\newcommand{\ea}{\end{eqnarray}}

\newcommand{\wmap}{{\it WMAP }}
\newcommand{\kel}{$^{\circ} K\hspace{1mm}$}
\newcommand{\hs}{\hspace{1mm}}
\newcommand{\vs}{\vspace{2mm}}
\newcommand{\lya}{Ly$\alpha \hspace{1mm}$}

\newcommand{\kms}{km s$^{-1}\hspace{1mm}$}

\begin{document}
\twocolumn[

\title{Ly$\alpha$ Radiation From Collapsing Protogalaxies I:\\Characteristics of the Emergent Spectrum.}

\author{Mark Dijkstra\altaffilmark{1,3}, Zolt\'an Haiman\altaffilmark{1} \& Marco Spaans\altaffilmark{2}}

\affil{$^1$Department of Astronomy, Columbia University, 550 West 120th Street, New York, NY 10027}

\affil{$^2$Kapteyn Astronomical Institute, P.O. Box 800, 9700 AV Groningen, The Netherlands}

\affil{$^3$School of Physics, University of Melbourne, Parkville, Victoria 3010, Australia}

\vspace{0.23cm}
\vspace{-0.5\baselineskip}
 
\begin{abstract}
We present Monte Carlo calculations of Ly$\alpha$ radiative transfer
through optically thick, spherically symmetric, collapsing gas clouds.
These represent simplified models of proto--galaxies that are caught
in the process of their assembly.  Such galaxies produce Ly$\alpha$
flux over an extended solid angle, either from a spatially extended
Ly$\alpha$ emissivity, or from scattering effects, or both.  We
present a detailed study of the effect of the gas distribution and
kinematics, and of the Ly$\alpha$ emissivity profile, on the emergent
spectrum and surface brightness distribution. The emergent Ly$\alpha$
spectrum is typically double--peaked and asymmetric. In practice,
however, we find that energy transfer from the infalling gas to the
Ly$\alpha$ photons -together with a reduced escape probability for photons
in the red wing- causes the blue peak to be significantly enhanced and the
red peak, in most cases, to be undetectable. This results in an
effective blueshift, which, combined with scattering in the
intergalactic medium, will render extended Ly$\alpha$ emission from
collapsing protogalaxies difficult to detect beyond redshift $z\gsim
4$. We find that scattering flattens the surface brightness profile in
clouds with large line center optical depths ($\tau_0>10^5$). A strong
wavelength dependence of the slope of the surface brightness
distribution (with preferential flattening at the red side of the
line) would be a robust indication that Ly$\alpha$ photons are being
generated (rather than just scattered) in a spatially extended,
collapsing region around the galaxy. We also find that for self--ionized
clouds whose effective Ly$\alpha$ optical depth is $\lsim 10^3$,
infall and outflow models can produce nearly identical spectra and
surface brightness distributions, and are difficult to distinguish
from one another. The presence of deuterium with a cosmic abundance
may produce a narrow but detectable dip in the spectra of systems with
moderate hydrogen column densities, in the range $10^{18}-10^{20}~{\rm
cm^{-2}}$. Finally, we present a new analytic solution for the
emerging Ly$\alpha$ spectrum in the limiting case of a static uniform
sphere, extending previous solutions for static plane--parallel slabs.
\end{abstract}

]


\section{Introduction}
\label{sec:intro}
The early stages of galaxy formation are characterized by a spatially
extended distribution of gas.  As gas cools and settles inside its
host dark matter halo, most of its gravitational binding energy is
radiated away. In dark matter halos with $T_{\rm vir}>10^4$ K, the
majority of this radiation can emerge in the Ly$\alpha$ line
\citep{Haiman00,Fardal01}.  Protogalactic gas clouds may therefore be
detected as spatially extended sources in the Ly$\alpha$ line.  In the
presence of ionizing sources enclosed by the collapsing gas, the total
Ly$\alpha$ flux from these clouds may be significantly boosted, as
photo-ionization and subsequent recombination converts the ionizing
continuum photons into Ly$\alpha$ radiation. The ionizing sources may
consist of young massive stars (which are hot enough to produce
substantial ionizing radiation) and/or quasars.  For example,
\citet{Haiman01} have shown that in the presence of a luminous central
quasar, this boost in the amount of Ly$\alpha$ emission from a
spatially extended protogalactic gas cloud may be several orders of
magnitude.

Based on the three Ly$\alpha$ emission mechanisms sketched above --
cooling radiation, or recombination powered by the ionizing radiation
of stars or quasar black holes, -- copious Ly$\alpha$ emission may be
expected from the early stages of galaxy formation,
during which the gas is spatially extended and collapsing.
 Simple estimates of the total Ly$\alpha$ line flux show that it can be detectable out to high redshifts with modern instruments. Indeed, Wide--field narrow--band surveys, such as for example the Large Area Lyman Alpha (LALA) survey \citep{rhoads00} have been very fruitful in finding young galaxies at redshifts $z=4.5-6.5$. Although the majority of the observed Ly$\alpha$ emitters appear
point--like (within the detection limits), several tens of extended
Ly$\alpha$ sources have currently been observed at $z\sim 3$
\citep[e.g][]{Steidel00,Bunker03,Matsuda04}. One may expect 
that a gas cloud that is caught in the early stages
 of galaxy assembly to be relatively metal-free, since not much star-formation
  has taken place. The observations of \citet{Weidinger04,Weidinger05} suggest
that such a pristine (not polluted by metals) gas cloud around a $z\sim3$
quasar has already been found. A similar extended Ly$\alpha$ "fuzz" has
been discovered around 4 additional quasars recently by
\citet{Christensen05}. \citet{Barkana03} have shown that evidence for
 gas collapse around quasars exists in the form of an absorption feature in the quasar's Ly$\alpha$ emission line. 

Future searches for both Ly$\alpha$ emitters and blobs will
undoubtedly reveal more Ly$\alpha$ emitters, point--like and extended.
Deeper observations and/or higher resolution spectra may contain
detailed information on the gas distribution and
kinematics. Therefore, it is timely to ask the following question:
{\it How can we use the Ly$\alpha$ radiation itself, to identify
these objects as collapsing protogalaxies-and to constrain the mechanisms
 powering their Ly$\alpha$ emission?}  

Quantifying the properties of Ly$\alpha$ 
radiation emerging from collapsing protogalaxies, such
as its expected surface brightness profile and spectral energy
distribution, is difficult because: (i) it is not well
understood how and where the Ly$\alpha$ photons are generated, (ii)
the optical depth $\tau$ in the line exceeds unity for columns of
neutral hydrogen $\gtrsim 10^{14}$ cm $^{-2}$, so that protogalaxies
are expected to be optically thick with $\tau\gsim 10^3$,
necessitating computationally expensive radiative transfer
calculations, and (iii) complex gas density and velocity distributions
\citep[e.g.][]{Kunth98}, as well as the presence of dust
\citep[e.g.][]{Neufeld91,Charlot93,Hansen06}, complicate any
simplified predictions.  Additionally, at redshifts $z\gsim 3$, the
intergalactic medium (IGM) is optically thick at the Ly$\alpha$
frequency, and will modify the observed spectrum.

In this paper, we attack this problem with Monte Carlo 
computations,  similar to those used in previous 
papers \citep[e.g.][]{Ahn00,Ahn02,Zheng02,Cantalupo05}
of Ly$\alpha$ transfer through a range of 
models that represent collapsing protogalactic gas clouds. 
 The main goal of this work is to gain basic insights
 into the properties of the emerging Ly$\alpha$ radiation.
 and into how these properties depend on the assumed
gas density distribution, gas kinematics and on where the Ly$\alpha$
is generated. Furthermore, we study how these predictions change in
the presence of an embedded ionizing source. 
Our focus is on the in the earliest 
stages of galaxy formation, when there is still
significant gas infall with an extended spatial distribution.
Therefore the geometry we study is that of a spherically symmetric,
collapsing gas cloud, which is obviously
oversimplified. However, the simple geometry is computationally less
expensive and allows us to explore a wide range of 
Ly$\alpha$ emissivity profiles and gas distributions
 that may be appropriate for the collapsing protogalaxies. 
Additionally, the absence of complicated geometrical
effects makes our results easier to interpret and may fulfill the same
role for future \citep[and existing, e.g.][]{Iro05}, 
more sophisticated 3-D simulations, that analytic
solutions have played for the present work. Furthermore,
we will argue that several of our result are unlikely to change
for more realistic, complex gas distributions and kinematics.
In a companion paper \citep[][hereafter Paper II]{PaperII}, 
we present a detailed comparison of our models with several 
existing observations of Ly$\alpha$ emitters.

The rest of this paper is organized as follows.
In \S~\ref{basics}, we briefly review the basics of Ly$\alpha$ transfer.
In \S~\ref{sec:codeandtest}, we describe our code, along with several tests we performed to check its accuracy. 
In \S~\ref{Deut}, we describe how deuterium was included in our 
calculations, and a motivation for why and when this is relevant.
In \S~\ref{sec:cooling}, we apply our code to Ly$\alpha$ transfer calculations of neutral collapsing gas clouds (without any embedded ionizing sources). 
In \S~\ref{sec:lyaresults}, we present present our main results on the emergent spectra and
surface brightness profiles.
In \S~\ref{sec:centralsource}, we investigate the additional effect of an embedded ionizing source.
In \S~\ref{sec:discuss}, we discuss 
the impact of the IGM and various caveats in our calculations. 
Furthermore, we compare our results with other work.
Finally,  in \S~\ref{sec:conclusions}, we present our conclusions and summarize the implications of this work. 
The parameters for the background cosmology used throughout this paper
are $\Omega_m=0.3$, $\Omega_{\Lambda}=0.7$, $\Omega_b=0.044$, $h=0.7$,
based on \citet{Spergel03}.

\section{Ly$\alpha$ Radiative Transfer Basics}
\label{basics}
The basics of transfer of Ly$\alpha$ resonance line radiation has been
the subject of research for many decades and is well understood
\citep[e.g.][]{Zanstra49,Unno52,Field59,Adams72,Harrington73,Neufeld90}.  The
goal of this section is to review the basics of Ly$\alpha$ scattering that is
relevant to understanding this paper. It is convenient to express
frequencies $\nu$ in terms of $x\equiv (\nu-\nu_0)/\Delta \nu_D$,
where $\Delta \nu_D=v_{th}\nu_0/c$, and $v_{th}$ is the thermal
velocity of the hydrogen atoms in the gas, given by $v_{th}=\sqrt{2k_B
T/m_p}$, where $k_B$ is the Boltzmann constant, $T$ the gas
temperature, $m_p$ the proton mass and $\nu_0$ is the central \lya
frequency, $\nu_0=2.47 \times 10^{15}$ Hz. 

The optical depth through a column of hydrogen, $N_{\rm HI}$,  
for a photon in the line center, $\tau_0$ is given by

\begin{eqnarray}
\tau_0&=N_{\rm HI} \hs \sigma_0=N_{\rm HI} 
\hs f_{12}\frac{\sqrt{\pi}e^2}{m_e c \Delta \nu_D}= \nonumber \\
&=8.3 \times 10^6 \Big{(}\frac{N_{\rm HI}}{2 
\times 
10^{20}\hs{\rm cm}^{-2}} \Big{)} \Big{(} \frac{T}{2 
\times 10^4}\Big{)}^{-0.5},
\label{eq:tau0}
\end{eqnarray} 
where $\sigma_0$ is the Ly$\alpha$ absorption cross section in the
line center, $f_{12}=0.4167$, the oscillator strength and $m_e$ and
$e$ are the mass and charge of the electron, respectively. The optical
depth for a photon at frequency $x$ is reduced to

\begin{equation}
\frac{\tau_x}{\tau_0}=H(a,x)=\frac{a}{\pi}\int_{-\infty}^{\infty}
\frac{e^{-y^2}dy}{(y-x)^2+a^2}=
\left\{ \begin{array}{ll}
         \ \sim  e^{-x^2}& \mbox{core};\\
         \ \sim \frac{a}{\sqrt{\pi}x^2}& \mbox{wing},\end{array} 
\right. 
\label{eq:phi}
\end{equation} 
where $a$ is the Voigt parameter and is the ratio of the Doppler to
natural line width, given by $a=A_{21}/4 \pi \Delta \nu_D$ $=4.7
\times 10^{-4}$ $(13 \hs {\rm km \hs s}^{-1}/v_{th})$, where $A_{21}$
is the Einstein A-coefficient for the transition.  The transition
between `wing' and `core' occurs at $x \sim 2.5-4.0$ for $a \sim
10^{-2}-10^{-6}$.

When a Ly$\alpha$ photon of frequency $x_i$ is absorbed by an atom, it
re--emits a Ly$\alpha$ photon of the same frequency in its own frame,
unless the atom is perturbed while in the excited $2p$ state. These
perturbations include collisions with an electron and photo-excitation
or ionization by another photon. In this case, the energy of the
re-emitted Ly$\alpha$ photon, $x_o$, is not equal to $x_i$.  In most
astrophysical conditions these perturbations can be neglected (see
appendix~\ref{sec:perturbations}), rendering the scattering coherent
in the frame of the atom.

Due to the atom's motion however, the scattering is not coherent in
the observer's frame. A simple relation exists between the incoming
and outgoing frequency, $x_i$ and $x_o$, and the velocity of the atom,
${\vec v_a}$:

\begin{equation}
x_o=x_i-\frac{{\vec v_a}\cdot {\vec k}_i}{v_{th}}+\frac{{\vec v_a} \cdot{\vec k}_o}{v_{th}}+
g({\vec k}_i \cdot {\vec k}_o -1)+\mathcal{O}\Big{(}\big{[}\frac{v_{th}}{c}\big{]}^2\Big{)},
\label{eq:xinxout}
\end{equation} %
where ${\vec k}_i$ and ${\vec k}_o$ are the propagation directions of
the incoming and outgoing photons, respectively.  The second to last
term on the RHS is the `recoil' term, which accounts for the transfer
of a small amount of momentum from the photon to the atom during each
scattering \cite[exactly as in Compton scattering,
e.g.][p.196]{RL79}. The factor $g$ is the average fractional amount of
energy transferred per scattering. \citet{Field59} showed that $g$ can
be written as $g=h \Delta \nu_D/2kT$ $= 2.6\times 10^{-4}$ $(
13\hs{\rm km \hs s}^{-1}/v_{\rm th})$. The effect of recoil is (and
has been verified to be) negligible in the applications presented in
this paper \citep{Adams71}.

Because $x_i$ and $x_o$ are related via eq.~(\ref{eq:xinxout}) this
case is referred to as `partially coherent'. The redistribution
function, $q(x_o,x_i)dx_o$, gives the probability that a photon that
was absorbed at frequency $x_i$ is re--emitted in the frequency range
$x_o \pm dx_o/2$ (in the observer's frame). Under the assumption that
the velocity distribution of atoms is (locally) Maxwellian,
\citet{Unno52} and \citet{Hummer62} calculated the redistribution
function analytically for partially coherent scattering (
their $q_{II}(x_o,x_i)$). Their calculation provides a good test
case for our code, which is discussed in Appendix~\ref{app:test}.

Photons in the core have a short mean free path, and the rate at which
they are scattered is high. The majority of these scattering events
leave the photon in the core, but a chance encounter with a rapidly
moving atom may move it to the wing, where its mean free path is much
larger.  In moderately optically thick media, $\tau_0 \lsim 10^5$, one
such encounter may be enough to enable the photon to escape in a
single flight. According to eq.~(\ref{eq:tau0}) this optical depth corresponds to a column density of $2.4 \times 10^{18}$ cm$^{-2}$
 for a gas at $T=2 \times 10^4$ K. This is approximately the column
density of a Lyman limit system. \citet{Adams72} demonstrated that in the case of extremely optically thick media, $\tau_0 \gtrsim 10^3/a$, photons do
not escape in a single flight, but rather in a single `excursion'.
During this excursion the photon scatters multiple times in the wing,
and diffuses through real and frequency space. Because of the
significantly increased mean free path, a photon predominantly
diffuses spatially, while it is in the wing.  Each scattering in the
wing pushes the photon back towards the core by an average amount of
$-1/x$ \citep{Osterbrock62}.

The transfer of Ly$\alpha$ photons through a static, uniform,
non-absorbing, plane--parallel scattering medium (hereafter referred
to as ``slab'') has been studied extensively. The spectrum of
Ly$\alpha$ photons emerging form both moderately and extremely
optically thick slabs is double peaked and, when bulk motions in the
gas are absent and when recoil is ignored, symmetric around the
origin.  The spectra of Ly$\alpha$ photons emerging from extremely
optically thick slabs have been calculated analytically. 
For the case in which  photons
are injected in the center of the slab, in the line center, the
location of the peaks at $\pm x_p$ is given by
$x_p=1.06(a\tau_0)^{1/3}$ \citep{Harrington73,Neufeld90}, in which
$\tau_0$ is the total line center optical depth from the center to the
edge of the slab
\footnote{Note that the value of $x_p$ quoted by
\citet{Harrington73} and \citet{Neufeld90}, is actually
$x_p=0.88(a\tau_0)^{1/3}$. This is due to their different definition
of $H(a,x)$, which is lower by a factor of $\sqrt{\pi}$ \citep{Ahn02}.}.

\section{The Code and Tests}
\label{sec:codeandtest}
\subsection{The Code}
\label{sec:code}
Our Monte Carlo computations are based on the method presented by
ZM02. Our code focuses on spherically symmetric gas configurations. We
use $N_{\rm grid}$ spherical shells, each with their own radius,
density, neutral fraction of hydrogen and velocity. The temperature is
assumed to be the same in each shell, although the code can easily be
adjusted to account for temperature variations.  Throughout this
paper, we used $N_{\rm grid}=1000$. Tests performed with $10$ times
more shells produced identical results. Runs performed with $\sim 10$
times less shells still produced visually indistinguishable
results. Because runs with $N_{\rm grid}=100$ barely reduced the
computing time, $N_{\rm grid}=1000$ was used. The shells were spaced
such that every shell contains roughly an equal column of hydrogen
from its inner to its outer radius.

The creation and transfer of individual Ly$\alpha$ photons is
described below. In the transfer problem, we identify three relevant
frames: the 'central observer' frame; the frame of the shell through
which the photon is traveling; and the frame of the atom that scatters
the Ly$\alpha$ photon (the frame of the atom differs from
that of the shell it is in, because of the atom's thermal motion).
 The `central observer' frame is the frame of an observer
located in the center of the sphere. The reason that the observer is
put in the center is that it fully exploits spherical symmetry. The
bulk motion of the gas is simply pointed toward (for infall) or away
(for outflows) from the observer. Photons propagating outwards and
inwards, propagate away and towards the observer, respectively. Form
now on the `central observer' frame will be referred to simply as the
observer's frame.

A quantity $`Q`$ measured in the frame of a shell, observer and the
atom is denoted by $Q'$, $Q$ and $\tilde{Q}$, respectively.  $R_n$,
with $n=1,2,3,...$, are random numbers generated between 0 and 1.

\vs 
{\it Step 1.} Given the emissivity per unit volume as a function of
radius, $j(r)$, we generate the radius of emission $r_{em}$ from the
probability distribution function:

\begin{equation}
R_1=\frac{\int_0^{r_{em}}r'^2 j(r')dr'}{\int_0^{r_{max}}r'^2 j(r')dr'}, 
\end{equation} 
where $r_{max}$ is the radius of the gas sphere's edge.  The location
of the photon in the sphere, ${\vec r} \equiv (p_x,p_y,p_z)$, is given
by $(p_x,p_y,p_z)=r_{em}$ $(S_2 \hs C_3$, $ S_2 \hs S_3$, $C_2)$,
where $C_n$ and $S_n$, with $n=1,2,3,...$, are $\cos 2 \pi R_n$ and
$\sin 2 \pi R_n$, respectively.
The emissivity $j(r)$ depends on what model we consider, which is
discussed in \S~\ref{sec:modelinput}.

\vs
{\it Step 2.} Given the location of emission, we generate the
direction of emission ${\vec k_i}=(k_x,k_y,k_z)$=$(R_4 \hs C_5$, $ R_4
\hs S_5$, $(1-R_4^2)^{0.5})$.  In the case the gas has a bulk
velocity, the frequency of the emitted photon is given by
$x_{i}=x'_{i}+{\vec k_i}\cdot {\vec v_{\rm bulk}}/v_{\rm th}$, where
${\vec v}_{\rm bulk}$ is the bulk velocity vector of the gas at the
location of emission. For simplicity, we set $x'_{i}=0$, which means
that the Ly$\alpha$ is initially always injected at the line center in
the frame of the gas shell in which it is emitted. In reality $x'_{i}$
is distributed according to the Voigt profile (eq.~\ref{eq:phi}).
This minor simplification does not affect our results at all. The
reason is that the Ly$\alpha$ photon is redistributed in frequency
each time it is scattered, rapidly erasing any `memory' of its initial
frequency. For our spherically symmetric models, ${\vec v}_{\rm
bulk}=v_{\rm bulk} \vec{r}/|\vec{r}|$, where $v_{\rm bulk}$ is the
infall/outflow speed of the gas at radius $r$. The sign of $v_{\rm
bulk}$ is negative/positive for infall/outflow, respectively.

\vs
{\it Step 3.} The optical depth $\tau$ the photon is allowed to travel
is given by $-\ln[R_6]$. To convert $\tau$ to a physical distance, we
calculate the physical distance $d_e$ to the edge of the shell in
which the photon is emitted in the direction ${\vec k_i}$.  The
corresponding optical depth to the edge of the shell is $\tau_e=d_e
\hs n_{H} \hs \sigma_0 \hs \phi(x'_{i})$, where $n_H$ is the number
density of neutral hydrogen atoms in the shell.  If $\tau_e > \tau$,
then the new location is given by ${\vec r}+\lambda {\vec k_i}$, where
$\lambda=d_e \hs \tau/\tau_e$.  If $\tau_e < \tau$ then, we obtain the
frequency of the photon in the frame of the next shell from
$x'_{i}=x_{i}-{\vec k_i}\cdot {\vec v_{\rm bulk}}/v_{\rm th}$, where
${\vec v_{\rm bulk}}$ is now the bulk velocity vector of the gas where
the photon penetrates the next shell. 
In the next shell $\tau$ is replaced by $\tau-\tau_e$. This process is
repeated until the photon has traveled the generated optical depth
$\tau$ to its new position ${\vec r}$.

\vs 
{\it Step 4.} When $|{\vec r}|>r_{max}$ , the photon's frequency and
the angle under which it passed through the surface of the sphere are
recorded. Otherwise, the photon is scattered by a hydrogen atom, in
which case we proceed to the next step.

\vs
{\it Step 5.} Once the location of scattering, ${\vec r}$, has been
determined, the total velocity of the atom, ${\vec v_a}$, that
scatters the Ly$\alpha$ photon is generated. The total velocity is
given by the sum of the bulk velocity of the gas plus the thermal
velocity: ${\vec v_a}={\vec v}_{\rm bulk}+{\vec v_{th}}$.  The thermal
velocity, ${\vec v_{th}}$, is
decomposed into two parts. Following ZM02, the magnitude of the atom's
velocity in the direction ${\vec k_i}$, $u_{||}$, is generated from
eq.~(\ref{eq:u3}), 
\begin{equation} u_{||}=\frac{a}{\pi
H(a,x'_i)}\int_{-\infty}^{R_9}\frac{e^{-y^2}}{(x'_{i}-y)^2+a^2} dy.
\label{eq:u3}
\end{equation}
This distribution reflects the strong preference for photons at
frequency $x'_i$, to be scattered by atoms to which they appear
exactly at resonance, $u_{II}=x'_i$. For large $x'_i$, however, the
number of these atoms reduces as $\propto e^{-x_i'^2}$ and the photon
is likely scattered by an atom to which it appears far in the wing.

The two (mutually orthogonal) components perpendicular to ${\vec
k_i}$, $u_{\perp 1,2}$, are drawn from a Gaussian distribution with
zero mean and standard deviation $2^{-1/2}\hs v_{th}$.
 Following the procedure described in Numerical
Recipes \citep{Press92}
\begin{eqnarray}
u_{\perp 1}&= \sqrt{-{\rm ln}[R_{11}]}\hs \cos(2 \pi R_{10}) \nonumber \\
u_{\perp 2}&= \sqrt{-{\rm ln}[R_{11}]}\hs \sin(2 \pi R_{10}),
\label{eq:slowcode}
\end{eqnarray} 
where $u_{\perp 1,2}$ is in units of $v_{\rm th}$. At this particular
step in the code, it is possible to speed up the code significantly,
which is described below (\S~\ref{sec:fastcode}).

\vs 
{\it Step 6.}  In the frame of the atom, the distribution of the
direction of the re-emitted photon with respect to that of the
incoming photon is given by a dipole distribution,

\begin{figure*}[t]
\vbox{ \centerline{\epsfig{file=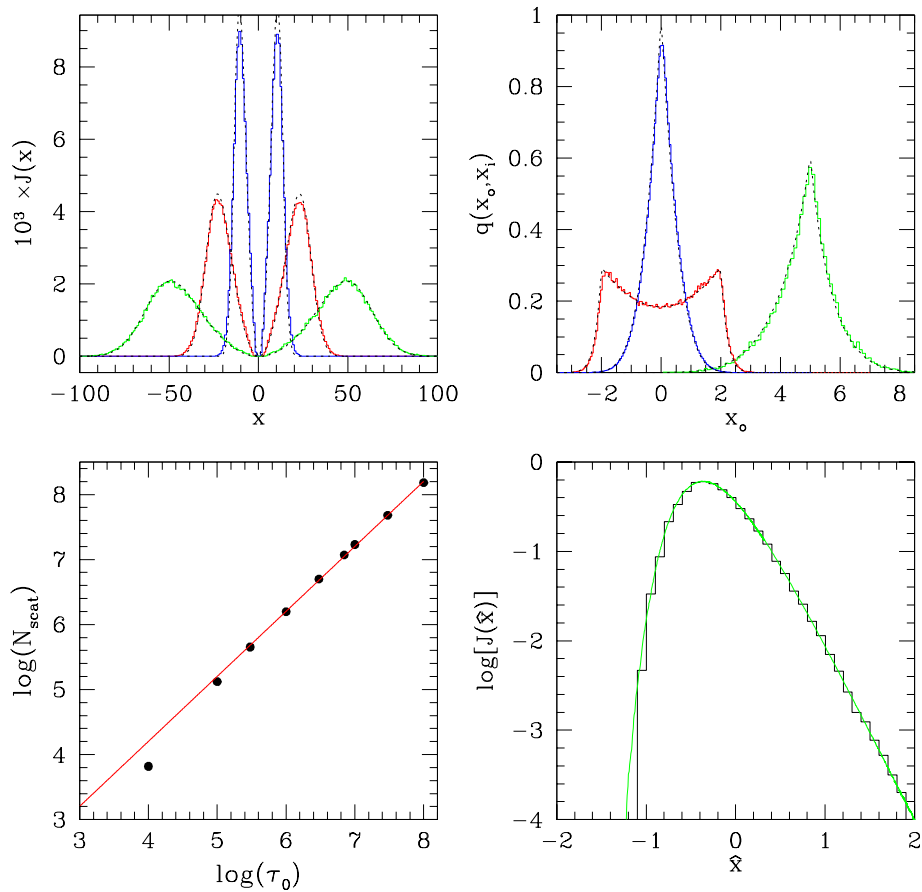,width=13.0truecm}}}
\caption[The tests performed on our \lya Monte Carlo code.]{ The
panels in this figure show four different tests of our Monte Carlo
code against known analytic solutions and Monte Carlo calculations by
\citet{RL99}. {\it Upper left panel:} The spectrum emerging from a
uniform spherical gas cloud, in which Ly$\alpha$ photons are injected
in the line center, $x=0$. The total line center optical depth,
$\tau_0$ from the center to the edge is $10^5$ and $10^7$ for the
narrowest ({\it blue}) and broadest ({\it green}) profile,
respectively. The intermediate spectrum ({\it red}) corresponds to
$\tau_0=10^6$. Overplotted as the {\it black dotted line} is the
theoretically derived spectrum, eq.~(\ref{eq:theoryspec}).  The tests
that are shown in the other panels are described in
Appendix~\ref{app:test}. {\it Upper right panel:} The colored
histograms are the redistribution functions, $q_{II}(x',x)$, for $x=0$
({\it blue}), $x=2$ ({\it red}) and $x=5$ ({\it green}), as extracted
from the code. The solid lines are the analytic solutions given in
\citet{Hummer62} and \citet{Lee74}.  {\it Lower left panel:} The total
number of scatterings a Ly$\alpha$ photon undergoes before it escapes
from a slab of optical thickness $2 \tau_0$ as extracted from the code
({\it  circles}). Overplotted as the {\it red--solid line} is
the theoretical prediction by \citet{Harrington73}. {\it Lower right
panel:}. The spectrum emerging from an infinitely large object that
undergoes Hubble expansion. The histogram is the output from our code,
while the {\it green--solid line} is the (slightly modified) solution
obtained by \citet{RL99} using their Monte Carlo algorithm. }
\label{fig:test} 
\end{figure*}
\begin{equation}
R_7 =\frac{3}{8}\int_{-1}^{\mu}(1+\mu'^2)d\mu' ,
\end{equation} 
where $\mu \equiv {\vec{k}_i} \cdot {\vec{k}_o}$.  Given
${\vec{k}_i}$, we have ${\vec{k}_o}=\mu {\vec{k}_i}+$
$(1-\mu^2){\vec{k}_{\perp}}$, where ${\vec{k}_{\perp}}$ is an
arbitrary unit vector perpendicular to ${\vec{k}_i}$. ${\vec k_o}$ is
calculated from ${\vec{k}_o}$ via a Lorentz transformation into the
observer's frame. Once ${\vec v_a}$ and ${\vec k_o}$ are known, $x_o$
can be calculated using eq.~\ref{eq:xinxout}. After replacing $x_i$
and ${\vec k_i}$ by $x_o$ and ${\vec k_o}$, respectively, we return to
step {\it 3}.

\subsubsection{Accelerated Scheme}
\label{sec:fastcode}

It is possible to speed up the code using a method described by
\citet{Ahn02}.  A critical frequency, $x_{\rm crit}$,
defines a transition from `core' to `wing'. Once a photon is in
the core ($|x'|<x_{crit}$), it is possible to force it (back) into the
wing, by only allowing the photon to scatter off a rapidly moving
atom. This is achieved by drawing $u_{\perp 1}$ 
and $u_{\perp 2}$ ({\it Step 5.} in \S~\ref{sec:code})
from a truncated Gaussian distribution, which is
Gaussian for $u > x_{\rm crit}$, but $0$ for $u < x_{\rm crit}$. To be
more precise:

\begin{eqnarray}
u_{\perp 1}&= \sqrt{x_{\rm crit}^2-{\rm ln}[R_{11}]}\hs \cos(2 \pi R_{10}) \nonumber \\
u_{\perp 2}&= \sqrt{x_{\rm crit}^2-{\rm ln}[R_{11}]}\hs \sin(2 \pi R_{10}),
\label{eq:fastcode}
\end{eqnarray} where $u_{\perp 1,2}$ is again in units of $v_{\rm th}$.

The reason it is allowed to skip core scattering is that
spatial diffusion occurs predominantly in the wings (see
\S~\ref{basics}). Skipping scattering events in the core
effectively sets the mean free path of core photons to 
$0$. Clearly, $x_{\rm crit}=0$ restores
the non--accelerated version of the code. Unless stated otherwise, in
this paper we used $x_{\rm crit}=3$. We found that larger values of
$x_{\rm crit}$ changed the appearance of the spectrum, especially the
presence of the red peak (see below) was strongly affected.

\subsection{The Tests}
\label{sec:test}

We performed various tests on the code, three of which are described
in Appendix~\ref{app:test}. The other test is explicitly described
here, since it involves a new analytic solution we derived in
Appendix~\ref{sec:analytic}. The test is similar to one generally done
by other groups \citep[e.g.][ZM02]{Ahn00,Ahn02,Cantalupo05}, but
differs in the geometry of the gas distribution. Ly$\alpha$ photons
are injected at $x=0$ in the center of a uniform, non--absorbing, 
static {\it sphere} with the gas at $T=10$ \kel, which corresponds to $a=1.5
\times 10^{-2}$. The line center optical depth from the center to the
edge of the sphere is $\tau_0$. In the {\it upper left panel} of
Figure~\ref{fig:test} the emergent spectra are shown for three
different values of $\tau_0$: $\tau_0=10^5,10^6$ and $10^7$ and
compared to the analytic solution given below.

\begin{equation}
J(x)=\frac{\sqrt{\pi}}{\sqrt{24}a\tau_0}\Bigg{(}\frac{x^2}
{1+{\rm cosh}\Big{[}\sqrt{\frac{2\pi^3}{27}}\frac{|x^3|}{a\tau_0}
\Big{]}}\Bigg{)}
\label{eq:theoryspec}
\end{equation}

The derivation of this analytic solution for a sphere is completely
analogous to the derivation given by \citet{Harrington73} and
\citet{Neufeld90}, and can be found in Appendix \ref{sec:analytic}. As
mentioned in \S~\ref{basics}, the emergent spectra are double--peaked
and symmetric around the origin, with $x_p$ increasing with
$\tau_0$. For a given value of $\tau_0$, $x_p=0.92(a\tau_0)^{1/3}$
(vs. $x_p= 1.06(a\tau_0)^{1/3}$ for a slab). The spectra are
normalized (as is the analytic solution) such that the area under all
curves is $(2\pi)^{-1}$. The figure shows that the agreement is
good. In Appendix~\ref{app:test} other tests are performed on the code
that involve core scattering and a case in which the gas has bulk
motions. Here, we briefly summarize these tests: In the {\it upper
right panel} the redistribution functions, $q_{II}(x',x)$, for $x=0$
({\it blue}), $x=2$ ({\it red}) and $x=5$ ({\it green}), as extracted
from the code are compared with the analytic solutions given in
\citet{Hummer62} and \citet{Lee74} ({\it solid lines}). The total
number of scatterings a Ly$\alpha$ photon undergoes before it escapes
from a slab of optical thickness $2 \tau_0$ as extracted from the code
({\it  circles}) is shown in the {\it lower left
panel}. Overplotted as the {\it red--solid line} is the theoretical
prediction by \citet{Harrington73}. In the {\it lower right panel} we
show the spectrum emerging from an infinitely large object that
undergoes Hubble expansion. The histogram is the output from our code,
while the {\it green--solid line} is the (slightly modified) solution
obtained by \citet[hereafter LR99]{RL99} using their Monte Carlo
code. As Figure~\ref{fig:test} shows, the code passes all these tests
well.

\section{Deuterium}
\label{Deut}
\begin{figure}[t]
\vbox{ \centerline{\epsfig{file=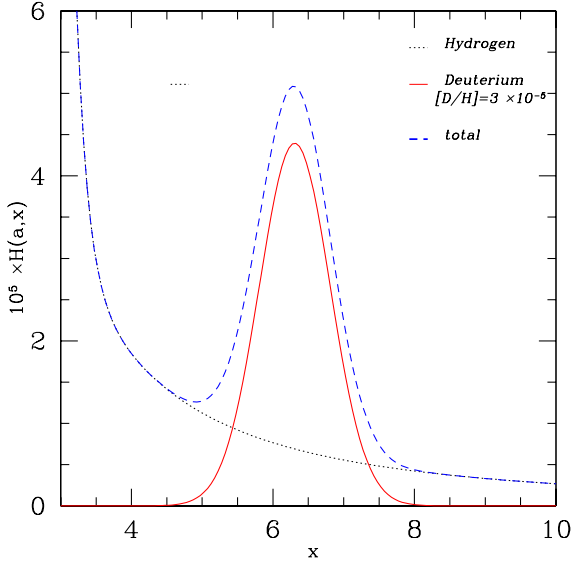,width=8.0truecm}}}
\caption[The contribution of deuterium to the total opacity.]{ The contribution of a cosmological abundance of deuterium ($[D/H]=3 \times 10^{-5}$) to the Ly$\alpha$ opacity for a gas at $T=10^4$ K. The ratio of the optical depth at frequency $x$ and frequency $x=0$, $\tau_{x,H}/\tau_0 \equiv H(a,x)$, through static gas is shown. The {\it black--dotted line} is for hydrogen only ( and is given by eq.~\ref{eq:phi}); the {\it red--solid line} is for deuterium only; the {\it blue--dashed line} is their sum. The figure shows that in the range $x=5-8$, deuterium dominates the opacity. For lower/higher temperatures, this spectral modification occurs at higher/lower values of $x$ with a relatively larger/smaller contribution from deuterium, respectively.}
\label{fig:voigt} 
\end{figure}

Although it has a low abundance, accounting for deuterium may be
relevant when predicting spectra emerging from Ly$\alpha$ emitters.
The resonance frequency of the deuterium Ly$\alpha$ absorption line is
blueshifted by $82$ \kms with respect to that of hydrogen.  For a gas
at $T \equiv T_4 \times 10^4$ K, this corresponds to $x_D \equiv
+6.3\hs (T_4)^{-1/2}$.  According to eq.~(\ref{eq:phi}), the hydrogen
optical depth at this frequency is reduced to $\tau_{x_D,H}\sim 6.6
\times 10^{-6}$ $\hs T_4^{1/2} \tau_0$.  At this frequency, the photon
is (by definition) exactly in the deuterium line center, and the
optical depth due to deuterium is $\tau_{x_D,D}=2^{1/2}[D/H]\tau_0$
$\sim 4.4\times10^{-5} \tau_0$, where $[D/H]$ is the number of
deuterium atoms per hydrogen atom\footnote{The factor $2^{1/2}$ reflects that deuterium is two
times heavier than hydrogen, which reduces its thermal velocity by
$2^{1/2}$, which according to eq.~(\ref{eq:tau0}) increases the line
center optical depth by $2^{1/2}$. Additionally, the reduced thermal
velocity of deuterium raises its Voigt parameter by $2^{1/2}$.}, $[D/H]=3 \times 10^{-5}$
\citep{Burles98}. The
contribution of deuterium to the optical depth, $\tau_{x,D}/\tau_0$,
is shown as a function of frequency in the range $x=3-10$ ({\it
red--solid line} in Figure~\ref{fig:voigt}).  The {\it black--dotted
line}) shows the original $\tau_{x,H}/\tau_0$ given by
eq.~(\ref{eq:phi}).  The ({\it blue--dashed line}) shows the sum of
the two. The gas is assumed to be at $10^4$ \kel. Clearly there is a
range in frequencies, $x\sim 5-8$, where deuterium dominates the
optical depth and regulates the transfer of Ly$\alpha$
photons. Despite extensive research on Ly$\alpha$ radiative transfer,
this detail has been neglected, to our knowledge, in all previous
studies.

To include deuterium in our code we added the contribution of
deuterium, $\tau_{x,D}$, to $\tau_x$. In addition to this, a step was
added to the code following {\it Step 4.} (\S~\ref{sec:code}), that
determines whether a scattering occurs by a hydrogen or deuterium
atom.  A random number $R_D$ was generated between 0 and 1. Let
$P_H\equiv\tau_{x,H}/[\tau_{x,H}+\tau_{x,D}]$ be the probability that
the Ly$\alpha$ photon is scattered by hydrogen.  When $R_D \leq P_H$,
the scattering is done by hydrogen, in which we proceed as described
under {\it Step 5.} in \S~\ref{sec:code}.  Otherwise, the scattering is
done by deuterium and its velocity vector is generated in exactly the
same way, but with $a$ and $x$ replaced by $2^{1/2}a$ and
$2^{1/2}[x-82\hs{\rm km}\hs{\rm s}^{-1}/v_{th}]$, respectively.

\begin{figure}[t]
\vbox{ \centerline{\epsfig{file=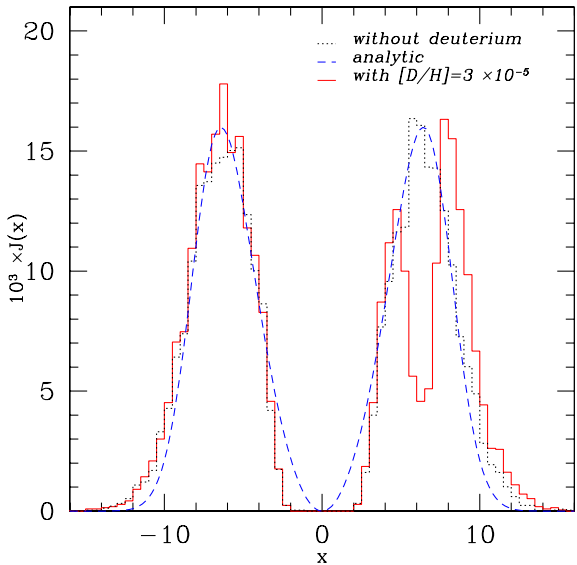,width=8.0truecm}}}
\caption[Example of imprint that deuterium can have on an emerging
\lya spectrum]{ The effect of a cosmological abundance of deuterium ($[D/H]=3 \times 10^{-5}$) on the spectrum emerging from a sphere with a center--to--edge column density of hydrogen of $N_{\rm HI}=1.2 \times 10^{19}$ cm$^{-2}$.  The reduced mean free path of Ly $\alpha$ photons in the deuterium line center inhibits their escape, which modifies the spectrum. For this plot, the gas temperature was assumed to be $10^4$ K. }
\label{fig:stat} 
\end{figure}

Figure~\ref{fig:stat} demonstrates the impact of deuterium on the
Ly$\alpha$ spectrum emerging from a uniform sphere in which the
photons are injected at line center ($x=0$) in the sphere's
center. The gas is at $T=10^4$ \kel and the total column of hydrogen
from the center to the edge of the sphere is $N_{HI}=1.2 \times
10^{19}$ cm$^{-2}$, yielding $\tau_0=7.3 \times 10^5$
(eq.~\ref{eq:tau0}). The emerging spectrum is shown with ({\it
red--solid--histogram}) and without ({\it black--dotted--histogram})
deuterium taken into account. The {\it blue--dashed line} is the
theoretically predicted spectrum, eq~(\ref{eq:theoryspec}). The reason
the theoretical curve does not fit very well is that the product
$a\tau_0=343$, while \citet{Neufeld90} requires $a\tau_0$ to be
$a\tau_0 \gtrsim 10^3$ for the theoretical curve to provide a good
fit. The imprint deuterium leaves on the spectrum is strong. The main
reason is that the presence of deuterium reduces the mean free path
for photons near $x=x_D$, which reduces the probability that they
escape the medium at this frequency. We find that even for a $10$
times lower deuterium abundance, the imprint in the spectrum is still
noticeable.

For deuterium to leave an observable imprint in the spectrum,
it should affect the frequencies at which a non-negligible fraction
of Ly$\alpha$ photons emerges (as in Fig.~\ref{fig:stat}). For $T_4=1$, for
example, we expect deuterium to leave a spectral feature for columns
in the range $N_{\rm HI}=10^{18}-10^{20}$ cm$^{-2}$, since this range
would translate to $x_p\sim(a\tau_0)^{1/3}\sim 3-12$. It is
interesting that this corresponds to the range of column densities in
Lyman limit systems. Because the columns studied in this paper are
much larger, typically $10^{21}-10^{22}$ cm$^{-2}$, the spectral
feature due to deuterium is much less prominent. Furthermore, bulk
motions in the gas will most likely smear the absorption feature over
a wide frequency range, which makes it even harder to observe in
practise. A further challenge is that the frequency range of the
deuterium feature corresponds to 3 thermal widths, or $v \sim 40$ km
s$^{-1}$. To resolve this narrow feature, spectra are required with
$R\sim 20000$. This resolution can be achieved with the High
Resolution Echelle Spectrometer (HIRES) on {\it Keck}. Its
detectability depends, however, on the total flux of Ly$\alpha$
emission. The brightest known Ly$\alpha$ blobs are $\sim 10^{-15}$
ergs s$^{-1}$ cm$^{-2}$ \citep{Steidel00}, which is $\lsim 2$
 orders of magnitude fainter than the sources successfully observed
 with this instrument. We plan to address the feasibility of detecting deuterium in more detail in future work.


\section{Ly$\alpha$ Emission from Neutral Collapsing Gas Clouds without Stars or Quasars}
\label{sec:cooling}
\subsection{Physics of Cloud Collapse and Ly$\alpha$ Generation}
\label{sec:coolingbasics}

For gas to collapse inside its host dark matter halo, it must lose its
gravitational binding energy.
 In the standard picture this binding energy is converted to kinetic
energy, which is converted to thermal energy when the gas
reaches the virial radius, where it is shocked to the virial
temperature of the dark matter halo, $T_{\rm vir}$. 
When the cooling time
is less than the dynamical time, the 
 gas practically cools down to $10^4$ \kel instantly.
Since this gas is no longer pressure supported, it falls inward on a
dynamical timescale \citep{Rees77}. 
Haiman et al. (2000) argued that a large fraction
of the cooling radiation is emitted in the Ly$\alpha$ line
(see also the early work by Katz \& Gunn 1991, who computed the
Ly$\alpha$ cooling radiation in a hydrodynamical simulation).
\citet{Birnboim03} presented a refined version of
the previous scenario, and used analytical arguments in combination
with a 1-D hydro simulation to show that virial shocks do not exist in
halos with $M \lsim 2 \times 10^{11}M_{\odot}$ (which at $z=3$
translates to $v_{\rm circ}\lsim 130$ km s$^{-1}$). Consequently, the
gas in these halos never gets shocked to the virial temperature and
the gas is accreted `cold' \citep[also
see][]{Thoul95}.

The details of cloud collapse and the associated Ly$\alpha$ emission are unclear. In \citet{Birnboim03}'s scenario, the gas falls unobstructed
into the center, where its kinetic energy is converted to heat, which
is radiated away in soft X-Rays. As a result, an HII region forms in which recombination converts $\sim68\%$ of the ionizing radiation into Ly$\alpha$. The details of the gas distribution in the center and on the mean energy of the ionizing photons determine the radius of this HII region, and thus how centrally concentrated the Ly$\alpha$ emitting region is. 


\citet{Haiman00} describe an alternative scenario that is supported by
3-D simulations, that show that any halo is build up of mergers of
pre-existing clumps. The denser colder clumps fall inwards, while experiencing weak secondary shocks from supersonic encounters with other density
inhomogeneities. As a result, the clumps are continuously heated,
which is balanced by cooling through Ly$\alpha$ emission. In the
 case the clumps fall in at a constant speed, all 
the gravitational binding energy of a clump is converted into 
heat which is radiated away in the Ly$\alpha$ line. The
Ly$\alpha$ emissivity associated with this scenario is spatially
extended. 

The main difference in the above models is the
 degree of concentration of the Ly$\alpha$ emissivity.
 While we cannot realistically  model the emissivity profiles ab-initio, 
we will below consider a wide range of profiles that should bracket 
the possibilities discussed above.

\subsection{Modeling the Gas Kinematics and Ly$\alpha$ Emissivity Profile}
\label{sec:modelinput}

Below, we define our model explicitly, and list the model parameters. We discuss the uncertainties of each parameter, and motivate the range of values 
we consider in each case. Input that is kept
fixed in all models is the density distribution of the dark matter,
$\rho_D(r)$, which is given by a NFW profile with concentration
parameter $c=5$. The gas temperature is assumed to be $10^4$ K
for all models. As discussed in \S~\ref{sec:discussion}, changing the
gas temperature does not affect our main results.  The density
distribution of the baryons, $\rho_b(r)$, is assumed to trace the dark
matter at large radii, with a thermal core at $r<3 R_s/4$
\citep[see][their eq. 9-11, where $R_s=R_{\rm
vir}/c$]{Maller04}. Although the results are not presented in this
paper, we found the effect of varying the concentration parameter $c$
to have no significant impact on our results. 

The total (dark matter + baryons) mass range of the
collapsing cloud, $M_{\rm tot}$, covered in this paper is $10^{10}-10^{13}
M_{\odot}$. A related quantity is the circular velocity, $v_{circ}$. The upper limit on this mass range is set by simple cooling arguments: the cooling time of objects exceeding $10^{13} M_{\odot}$ exceeds the Hubble time
\citep[e.g][]{Blumenthal84}. The lower limit corresponds to the mass
that can collapse even in the presence of an ionizing background \citep[$10^{10} M_{\odot}$ corresponds to $v_{\rm circ}\sim 50$ km s$^{-1}$ at $z=3$, see
e.g.][]{Thoul96,Kitayama00,Dijkstra04}. The redshift range at which the system virializes, $z_{\rm vir}$ covered in this paper is $z_{\rm vir}=3-6$. The lower limit is set by the redshift of the observed Ly$\alpha$ blobs which are candidates for cooling radiation. The upper limit is set by the opacity of the IGM which drastically increases beyond $z \gtrsim 6$ \citep[e.g.][]{Fan02}. Other input that is varied includes:\\

The velocity field, $v_{\rm bulk}(r)$, is given by a simple
power law ${\rm v}_{\rm bulk}(r)=v_{\rm amp}[r/r_{vir}]^{\alpha}$.  
The velocity profile is uncertain; it can conceivably increase or 
decrease with radius.
A spherical top-hat model would have $\alpha=1$ and $v_{\rm amp}=v_{\rm
circ}$.\footnote{This can be seen by time--reversing solutions for
trajectories initially following Hubble expansion, $v \propto r$.} In
more realistic initial density profiles the mean density 
within radius $r$ decreases smoothly with radius, 
causing the inner shells to be decelerated more relative to the 
overall expansion of the
universe. This enhances the infall speed at small $r$ relative to that
in the top-hat model.  The result of this is a flatter velocity
profile, i.e $\alpha < 1$. Accretion of massless shells onto a 
point mass results in $\alpha=-1/2$ \citep{Bertschinger85}.
Since it spans the range we can motivate physically,
we study values of $\alpha$ in the range $\alpha \in
[-0.5,1]$, but note $\alpha$ may in fact be $\lsim 0.5$.  
For cases with $\alpha<0$, the velocity diverges as $r
\rightarrow 0$. This artificially boosts the emission at small $r$. To
prevent this, these cases are modified to flatten at $r \lesssim r_v$:

\begin{equation}
v_{\rm bulk}(r)=v_{\rm amp}\Big{[}\frac{r_{v}+r}{r_{v}+r_{\rm vir}}\Big{]}^{\alpha}. 
\label{eq:vfield}
\end{equation} 
When $r \gg r_{v}$, $v_{\rm bulk}=v_{\rm amp}[r/r_{\rm vir}]^{\alpha}$
and when $r \rightarrow 0$, then $v_{\rm bulk} \rightarrow v_{\rm
max}$.  We set $v_{\rm max}=2v_{\rm amp}$, which uniquely determines
$r_v$.  The amplitude of the velocity field $v_{\rm amp}$ is also
varied between $0$ and $2$ $v_{\rm circ}$.\\

The emissivity, $j(r)$, which we take to be either `extended'
or `central. The extended case refers to the scenario sketched by
\citet{Haiman00}, in which the gas continuously emits a fraction
$f_{\alpha}$ of the change of its gravitational binding energy in the
Ly$\alpha$ line. The emissivity as a function of radius $j(r)$, in
ergs s$^{-1}$ cm$^{-3}$, is then given by

\begin{equation}
j(r)=f_{\alpha}\frac{G \hs M_{\rm tot}(<r)\hs \rho_b}{r^2} \frac{dr}{dt},
\label{eq:jr}
\end{equation} 
where $M_{\rm tot}(<r)$ is the total mass enclosed within radius $r$,
$\rho_{b}$ the mass density of baryons at radius $r$ and $dr/dt$ is
simply $v_{bulk}$, the speed at which the gas at radius $r$ is falling
in. We assume $f_{\alpha}$=1, which reflects extreme cases in which
all gravitational binding energy is converted into Ly$\alpha$. The
results given below scale linearly with $f_{\alpha}$.\\ The `central',
$j(r)=\mathcal{K}\delta(r)$, models represent the extreme version of
the scenario sketched by \citet{Birnboim03}, in which all \lya is only
generated in the center. Alternatively, these central models may represent any model
in which a central \lya emitting source is surrounded by a neutral
collapsing gas cloud.  We determine $\mathcal{K}$ from we the total
Ly$\alpha$ luminosity, given by $L_{{\rm ly} \alpha} \equiv \int \hs
dV j(r)$. For simplicity, we set $L_{{\rm ly} \alpha}$ equal in the extended
  and central models with otherwise identical parameters.\\

A brief summary of all models studied in this paper is given in
Table~\ref{table:models} in Appendix~\ref{app:table}. All results below are obtained by setting $r_{\rm max}=r_{\rm vir}$ (\S~\ref{sec:code}), which implies that we ignore the Ly$\alpha$ transfer beyond the virial radius. The impact of the IGM surrounding the virialized halo can vary from location to location (e.g.\ due to stochastic density and ionizing background fluctuations and also from the proximity effect near bright QSOs). 
We take the approach of studying the effects of scattering in the virialized halo and in the IGM separately. In this paper, we only compute the Ly$\alpha$ spectra as they emerge at the surface of the halos at $r_{\rm vir}$ with a brief discussion of the effect of the IGM in \S~\ref{sec:IGMI}. the IGM is discussed in more detail in PaperII.

\section{Results}
\label{sec:lyaresults}

Before discussing the spectral features and surface brightness
profiles, it is useful to express the total Ly$\alpha$ luminosity,
$L_{{\rm Ly}\alpha}$ ($L_{{\rm ly} \alpha} \equiv \int \hs
dV j(r)$, \S~\ref{sec:modelinput}), as a function of our model parameters:

\begin{equation}
\frac{L_{{\rm Ly}\alpha}}{10^{42}}= 0.66\hs \Big{(} 
\frac{M_{\rm tot}}{10^{11}}\Big{)}^{\frac{5}{3}} 
\Big{(} \frac{v_{amp}}{v_{\rm circ}}\Big{)}
\Big{(}\frac{1+z_{\rm vir}}{5}\Big{)}^{\frac{5}{2}} 
\Big{(} \frac{2-\alpha}{1.75}\Big{)}^{1.2} \hs \frac{{\rm ergs}}{{\rm sec}}.
\label{eq:lum}
\end{equation} 
The scalings of $L_{{\rm ly}\alpha}$ with $M_{\rm tot}$ and $v_{amp}$
and $(1+z_{\rm vir})$ are exact. The $\alpha$ dependence is a fit, that
is accurate to within $10\%$, over the range $\alpha \in
[-0.5,1]$. Note that equation~(\ref{eq:lum}) can be written as:

\begin{equation}
\frac{L_{{\rm Ly}\alpha}}{10^{42}}= 0.66\hs \Big{(} \frac{{\rm v}_
{\rm circ}}{116 \hs {\rm km}\hs{\rm s}^{-1}}\Big{)}^{5} 
\Big{(} \frac{v_{amp}}{v_{\rm circ}}\Big{)}
\Big{(} \frac{2-\alpha}{1.75}\Big{)}^{1.2} \hs \frac{{\rm ergs}}{{\rm sec}}.
\label{eq:lum2}
\end{equation}

The dependence of $L_{{\rm Ly}\alpha}$ on $v_{\rm circ}$ can be
understood from the following scaling relations: $L_{{\rm
Ly}\alpha}\propto U/t$ ($U$ is the total 
gravitational binding energy and $t$ is the timescale over
 which it is released, which is taken to be the dynamical timescale of the halo)
$\propto U/(R_{\rm vir}/v_{\rm circ})$ $=
v_{\rm circ} U/R_{\rm vir}$ $\propto v_{circ} M^2/R_{\rm vir}^2$
$\propto v^5_{\rm circ}$, where we used that $U \propto GM^2/R_{\rm
vir}$ and $v^2_{\rm circ}=GM/R_{\rm vir}$.

\begin{figure*}[ht]
\vbox{ \centerline{
\epsfig{file=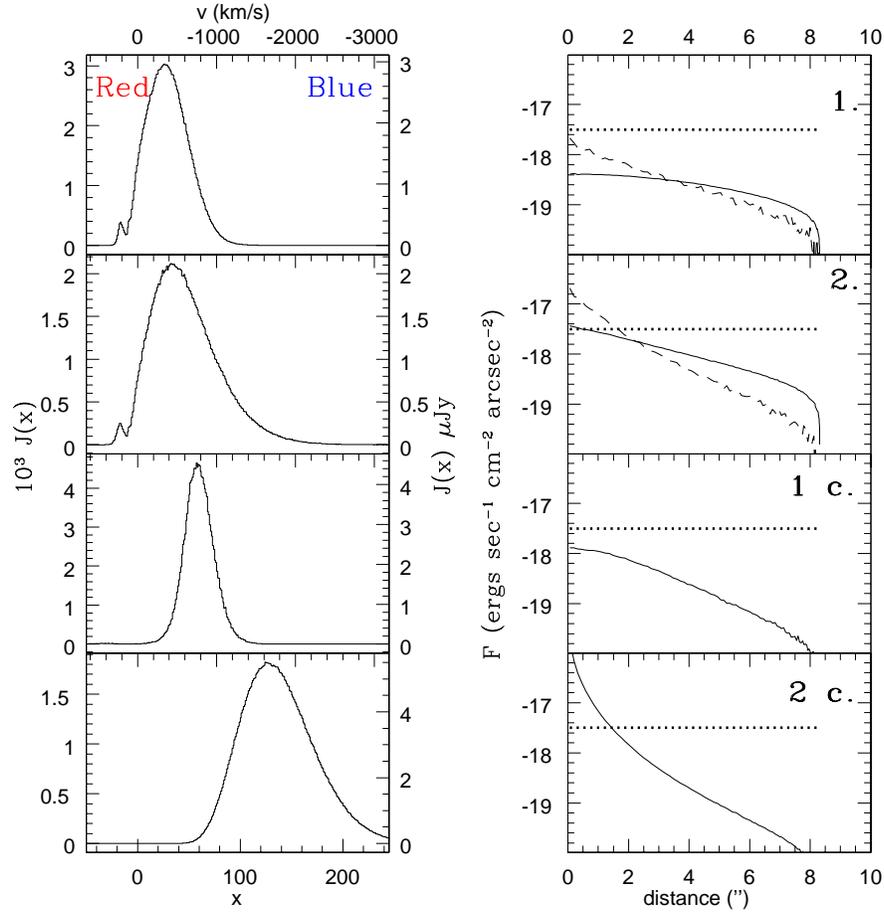,width=12.0truecm}}
\figcaption[\lya characteristics emerging from four models]{ Properties 
of the Ly$\alpha$ emission emerging from our four fiducial
models (models {\it 1., 2, 1 c} and {\it 2 c}) 
of neutral collapsing gas clouds.  The {\it left panels} show spectra in which the lower  horizontal axis shows the normalized frequency variable $x$. The upper horizontal axis shows what velocity offset this frequency corresponds
to. Values of $x>0$ correspond to higher frequencies, or lower
redshifts, which corresponds to lower recessional velocities. The left
 vertical axis shows the normalized flux, while the right vertical 
axis has the flux density in $\mu$Jy. Processing by the IGM is ignored in the plot. The {\it right panels} show the corresponding surface brightness distribution. The {\it
thick dotted horizontal line} corresponds to the detection limit in a
recent survey for Ly$\alpha$ blobs by \citet{Matsuda04}. The model
number is shown in the top right corner. The details of each model can
be found in Table~\ref{table:models}. The {\it black--dashed line}, in
the {\it upper two panels} in the right column, is the surface
brightness profile that would emerge, in the absence of
scattering. For the central models, this line corresponds to a
$\delta-$function at $0$'', and is therefore not shown. For all infall models, the Ly$\alpha$ line emerges with a net blueshift.
\label{fig:panel1}}}
\end{figure*}

In \S~\ref{sec:fiducialmodel} we show some examples of Ly$\alpha$ spectra and surface brightness profiles emerging from several models.
The main purpose of this is to gain basic insight into the Ly$\alpha$ radiation emerging from neutral, collapsing gas clouds with no star formation or quasar activity. In \S~\ref{sec:closerlook} the dependence of these properties on the model parameters is studied in more detail. Our main findings are summarized and and qualitatively explained in \S~\ref{sec:summarylya}.

\subsection{Examples of Intrinsic Spectra and Surface Brightness Profiles}
\label{sec:fiducialmodel}
We show the emerging spectrum ({\it left panels}) and surface brightness profile ({\it right panels}) for models {\it 1., 1 c, 2} and {\it 2 c.} in Figure~\ref{fig:panel1}. These models cover the full range of velocity profiles and Ly$\alpha$ emissivities. The model number associated with a given row shown in the top right corner of the {\it right panels}.
  The lower horizontal axis on the left column denotes the usual frequency variable $x$, which is converted to a velocity offset (in the
Ly$\alpha$ emitter's rest--frame) from the line center, on the upper
axis. Note that positive values of $x$ correspond to higher energies,
or bluer photons. The vertical axis on the left
side has units which are such that the area under the curves
corresponds to $(2 \pi)^{-1}$ ( as in Fig.~\ref{fig:test}). The
vertical axis on the right indicates the physical flux density in units of
$\mu$ Jy (ignoring scattering in the IGM, see \S~\ref{sec:IGMI}). 
The {\it right panels} panels show the surface brightness profiles.
The horizontal axis denotes the impact parameter, measured
in $''$. The units of the surface brightness on the vertical axis are
ergs s$^{-1}$ cm$^{-2}$ arcsec$^{-2}$. The thick dotted horizontal
line is the detection limit in the recent survey by \citet{Matsuda04}.

First, we focus on the spectra which show that the \lya radiation
escapes the collapsing protogalaxy predominantly blue shifted for all
models. The extended models are not as blue as the central models, and they
contain a faint red peak. We quantify the prominence of the red peak through 
the ratio $[B]/[R]$, defined as the number of photons in the blue peak over the number of photons in the red peak. 

Second, we focus on the surface brightness distribution.  Model {\it
1.} has a very flat profile, $F(\theta) \propto \theta^{-a}$, with $a \sim
0.5$. Model {\it 2.} has a steeper profile with $a \sim
1.0$ and the central surface brightness is increased above \citet{Matsuda04}'s detection limit. The profile that would emerge in the absence of scattering is
overplotted as the {\it dashed curves}. These curves illustrate the extent 
to which the surface brightness distribution becomes shallower 
because of scattering. The profiles for models {\it
1 c.} and {\it 2 c.} are more centrally concentrated ($a \sim 2$), which
can be viewed simply as an extension of the trend seen in the profiles
of models {\it 1.} and {\it 2.}: A steeper emissivity profile
translates to a steeper surface brightness profile.

\subsection{Impact of Model Parameters on the Emitted Ly$\alpha$ Features}
\label{sec:closerlook}

Here we discuss the impact of varying the model parameters described in \S~\ref{sec:modelinput} 
on on the following properties of the emergent spectrum: (i) net blueshift of the blue peak, (ii) 
FWHM of the blue peak, (iii)  the location of the red peak (if any), (iv) the ratio
$[B]/[R]$, and (v) the surface brightness profiles.

\subsubsection{Spectral Shape: Peak Morphology-The Blue Peak}
\label{sec:bluepeak}
\begin{figure}[ht]
\vbox{ \centerline{\epsfig{file=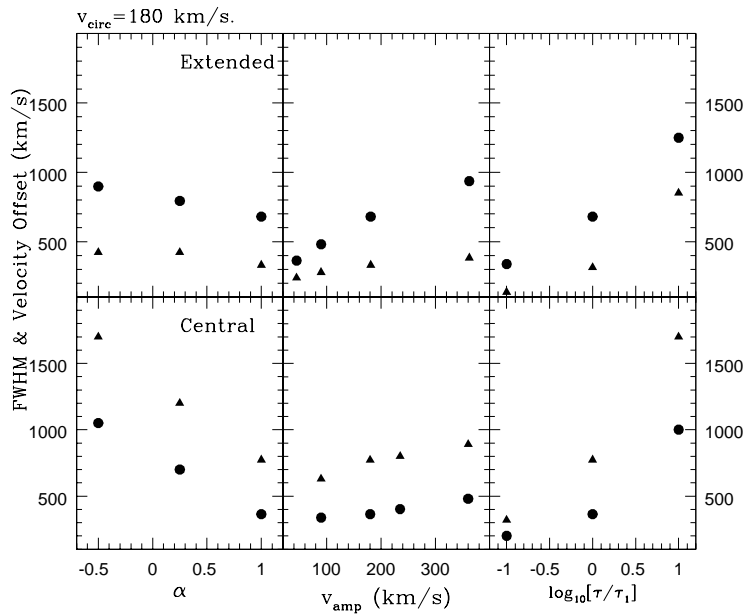,width=10.0truecm}}}
\caption[ The location in frequency of the maximum of the blue peak
and its FWHM for various models.]{ This figure shows the morphology of
the blue peak as a function of various model parameters. The location
of the maximum of the blue peak ({\it triangles}) and its FWHM
({\it circles}) as a function of 3 parameters for the
extended ({\it upper panel}) and central ({\it lower panels})
models. The {\it left panel}, {\it middle panel} and {\it right panel}
shows the dependence on $\alpha$, $v_{\rm amp}$ and the optical depth
$\tau$. The optical depth $\tau$ is normalized to $\tau_1$, which is
the optical depth from center to edge of model {\it 1.} (see
text). These trends suggest degeneracies will exist between the quantities $\tau$, $\alpha$ and $v_{\rm amp}$ - for example, larger infall velocities can be mimicked by larger column densities and vice versa (and result in a larger blueshift of the Ly$\alpha$ line).}
\label{fig:fwhm} 
\end{figure}
In Figure~\ref{fig:fwhm}. we show the frequency offset of the blue peak's 
maximum ({\it triangles}) and its FWHM ({\it 
circles}) as a function of three
parameters: $\alpha$ ({\it left panels}), $v_{\rm amp}$ (
{\it middle panels}) and the optical depth $\tau$ from
the center to the edge of the object ({\it right panels}). 
The upper and lower row of panels represents extended 
and central models, respectively.

The {\it left panels} show that for the extended
models, the dependence of the location of the maximum on
$\alpha$ is weak, while its FWHM is slightly more affected by
$\alpha$ (which is not surprising, as we notice that the blue tail
is more pronounced in the spectrum of model {\it 2.} than in model {\it 1.} in Fig.~\ref{fig:panel1}).  On the other hand, the dependence on $\alpha$ is very strong for the central cases (which is also evident from the lower two rows in
Figure~\ref{fig:panel1}).\\
 The {\it central panels} show that for the extended
 models the FWHM depends more strongly on $v_{\rm amp}$ than 
the peak's location. For the central models, the dependence 
of both quantities on $v_{\rm amp}$ is weaker than their dependence on $\alpha$. Also, the dependence of the FWHM on $v_{\rm amp}$ for the central
models is slightly weaker relative to that of their extended
counterparts.\\
 The {\it right panels} of Figure~\ref{fig:fwhm} shows the location of the
maximum of the blue peak and its FWHM for three different optical
depths, $\tau$.  The optical depth is normalized to that of model {\it
1.}, denoted by $\tau_1$.  All models are identical in their dark
matter and gas properties, but the Ly$\alpha$ absorption cross section
is decreased and increased by a factor of $10$ for the point with
$\tau/\tau_1$ $=0.1$ and $10.0$, respectively. The figure demonstrates
clearly that the blue peak's morphology is strongly affected by the
value of $\tau$. Both the FWHM and velocity shift of the Ly$\alpha$ line increase with $\tau$. These trends suggest that degeneracies will exist between the values of $\tau$, $v_{\rm amp}$ and $\alpha$ - for example, larger infall velocities can be mimicked by larger column densities and vice versa.

\subsubsection{Spectral Shape: Peak Morphology-The Red Peak}
\label{sec:redpeak}
\begin{figure}[ht]
\vbox{ \centerline{\epsfig{file=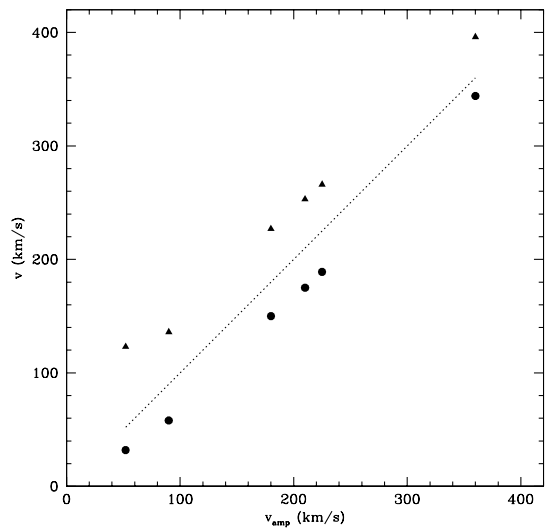,width=8.0truecm}}}
\caption[The location in frequency of the red peak for various
models.]{The location of the maximum of the red peak ({\it
-circles}) and the minimum in the spectrum ({\it
-triangles}) as a function of $v_{\rm amp}$. The {\it dotted
line} corresponds to the curve $v=v_{\rm amp}$. Despite the fact that
the included models vary in mass, redshift and $\alpha$, the correlation is obvious, which suggests it is a generic property of the extended models. The location of the red peak potentially measures the gas infall speed.}
\label{fig:redpeak} 
\end{figure}
 Figure~\ref{fig:redpeak} shows the location of the dip ({\it circles}) and the maximum of the red peak ({\it triangles}) as a function of the amplitude of the velocity field for various models that vary in mass, redshift and their assumed velocity field. The {\it dotted line} denotes $v=v_{\rm amp}$. The exact parameters used for these models used in the plot can be read off from Table~\ref{table:models}.

\begin{figure}[b]
\vbox{ \centerline{\epsfig{file=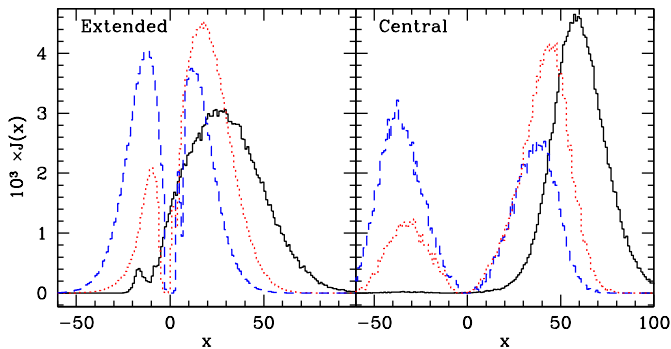,width=10.0truecm}}}
\caption[Effect of bulk motion on the emerging \lya spectrum.]{The
effect of bulk motions in the gas on the emerging Ly$\alpha$ spectrum
is shown for the extended ({\it left panel}) and central ({\it right
panel}) cases. The panels contain normalized spectra for three models
that only differ in the amplitude of the velocity field. The {\it
black--solid line} in the left/right panel corresponds to model {\it
1.}/{\it 1 c.}, respectively. The {\it red dotted lines} and 
{\it blue dashed line} correspond to the same models with 
$v_{\rm amp}$ reduced to $0.25v_{\rm circ}$ and {\it 15 c.}) and $v_{\rm amp}=0$, respectively. The red peak becomes smaller with increasing infall speed.}
\label{fig:sequence} 
\end{figure}

The relation between $v_{\rm amp}$ and the location of the peak and
dip is evident. Beyond $v_{\rm amp}\sim 360$ km s$^{-1}$ 
the correlation breaks down.  The reason is that the red
and the blue peak merge for large $v_{\rm amp}$ (see Fig~\ref{fig:fwhm} and Fig~\ref{fig:sequence} and their discussion below). 
The correlation holds for a wide range of models, 
which we take as evidence for it being a generic property of 
the extended models. 

\begin{figure}[htb]
\vbox{ \centerline{\epsfig{file=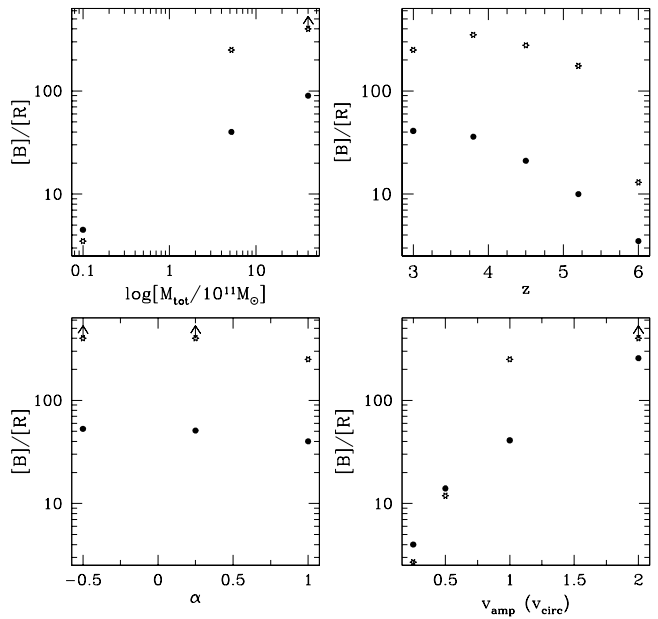,width=9.0truecm}}}
\caption[The ratio blue to red ratio as a function of $\alpha$,
$z_{\rm vir}$, $v_{\rm amp}$ and $M_{\rm tot}$.]{ The ratio of the
number of photons in the blue and red
peak, $[B]/[R]$, as a function of various parameters. The fiducial
model is model {\it 1.} (see Table~\ref{table:models}). Variation
with: {\it upper left panel:} mass, {\it upper right panel:} 
redshift (In this panel only, we include the scattering in the IGM, 
since it introduces a strong redshift dependence, see \S~\ref{sec:IGMI}), {\it lower right panel:} $v_{\rm amp}$,
{\it lower left panel:} $\alpha$. The red peak's prominence is enhanced towards higher redshifts, lower masses and lower infall velocities.}
\label{fig:boverr} 
\end{figure}
As mentioned in \S~\ref{sec:fiducialmodel} the
blue to red ratio, $[B]/[R]$ quantifies the prominence of the red peak.
First we demonstrate how the transition occurs from a double peaked
symmetric spectrum, known from the static cases, to the asymmetric
spectra, that are associated with bulk motions. The {\it left} and
{\it right panel} of Figure~\ref{fig:sequence} contains the
(normalized) spectrum of models {\it 1.}  and {\it 1 c.}, respectively,
as the {\it black--solid lines.} The {\it red--dotted lines}
correspond to the same models, in which the amplitude of the infall
velocity has been reduced to $0.25v_{\rm circ}$. The {\it blue dashed
lines} correspond to $v_{\rm amp}=0$ (actually, $v_{\rm
amp}=10^{-6}v_{\rm circ}$, because for $v_{\rm amp}=0$ there is no
emission for the extended cases according to eq.~\ref{eq:jr}).

The slight asymmetry that is present in both static cases, is
intriguingly enough, due to deuterium. Deuterium also leaves a sharp
feature at $x=x_D$ in the static, extended, case. The effect of
deuterium on the other spectra was found to be
negligible. Figure~\ref{fig:sequence} shows that $[B]/[R]$ increases
strongly with $v_{\rm amp}$ for both the central and extended
models. In the central cases, the blue and red peak remain well
defined and separated as the amplitude of the infall velocity is
increased. However, the amplitude of the red peak falls quickly to be undetectable. For the extended cases, the FWHM of the 
blue peak increases, and that of
the red peak decreases, until the peaks merge. The minimum in the spectrum
 that separates the blue and the red peak moves
to redder frequencies with increasing $v_{\rm amp}$.

Figure~\ref{fig:boverr} shows the value of $[B]/[R]$ as a
function of $z_{\rm vir}$ ({\it upper-right panel}),  
$v_{\rm amp}$ ({\it lower-right panel}),  
$\alpha$ ({\it lower-left panel}) and $M_{\rm
tot}$ ({\it upper-left panel}).
 The {\it  circles} refer to the extended models,
whereas the {\it open stars} refer to the central models.\\ 
The {\it upper-left panel} shows that $[B]/[R]$ increases strongly
with mass. 
The {\it upper-right panel} shows that $[B]/[R]$ decreases
strongly with  $z_{\rm vir}$. In this panel only, 
we included the impact of the IGM, because the 
IGM introduces a strong z-dependence. A discussion of how the IGM was
incorporated is delayed to \S~\ref{sec:IGMI}.
The {\it lower-right panel} shows that the value of $[B]/[R]$ 
increases with the amplitude of the velocity field $v_{\rm amp}$.
For sufficiently large $v_{\rm amp}$, 
the red peak is absorbed in the blue peak (see Fig~\ref{fig:fwhm}),
 in which case $[B]/[R] \rightarrow \infty$.
The {\it lower-left panel} shows that for the
extended models $[B]/[R]$ is relatively insensitive to the value of
$\alpha$. For the central models,  no photons emerged on the red 
side of the line for $\alpha \leq 0.25$, 
causing $[B]/[R] \rightarrow \infty$.\\
These results suggest that the secondary red peak may be 
detectable in the spectra of (proto) galaxies, provided the Ly$\alpha$ emissivity is spatially extended. Furthermore, the red peak may be more easily detectable towards higher redshift. If detected, the location of the red peak
 gives a potentially useful measure of the gas infall speeds.

\subsubsection{The Surface Brightness Profile}
\label{sec:surfacebrightness}
\begin{figure}[ht]
\vbox{
\centerline{\epsfig{file=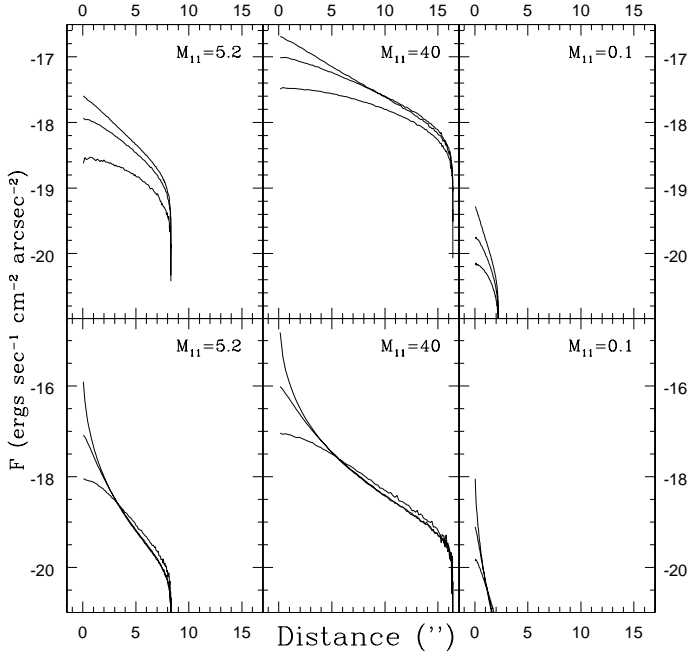,width=9.5truecm}}}
\caption[The surface brightness profiles for a set of models]{The
surface brightness as a function of angular distance in $''$ from the
center of the Ly$\alpha$ image. The {\it upper} and {\it lower panels}
represent extended and central models, respectively. Each panel
corresponds to a different mass, which is given in units of
$10^{11}M_{\odot}$ in the upper right corner of each panel.
The velocity profiles have slopes $\alpha=$ $-0.5$ (steepest curve), $0.25$ and $1.0$ (shallowest curve). The figure shows that the surface brightness profile becomes flatter with increasing mass. For a given mass, the exact slope is determined by how centrally concentrated the Ly$\alpha$ emissivity is. }
\label{fig:surbrightness} 
\end{figure}

Finally, we focus on the impact of varying the input parameters on the
emerging surface brightness profile. Figure~\ref{fig:surbrightness}
shows the surface brightness distribution as a function of $M_{\rm
tot}$, $\alpha$ and $j(r)$.  The {\it upper and lower panels} are
associated with extended and central models, respectively. Each
panel contains three models for one particular mass, $M_{\rm tot}$,
which is given in the top right corner in units of
$10^{11}M_{\odot}$. In all panels, the lowermost and uppermost curve
corresponds to $\alpha=1.0$ and $-0.5$, respectively (also see
Fig.~\ref{fig:panel1}), the middle curve has
$\alpha=0.25$. Figure~\ref{fig:surbrightness} shows that the surface
brightness profile flattens with increasing mass and $\alpha$.

\begin{figure}[ht]
\vbox{ \centerline{\epsfig{file=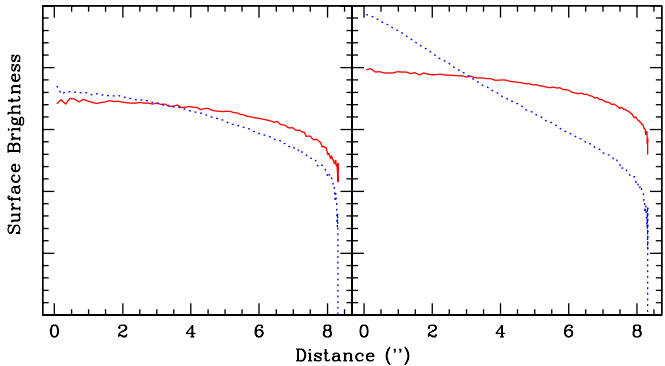,width=10.0truecm}}}
\caption[The surface brightness profile for the reddest and bluest
photons.]{This figure shows the frequency dependence of the surface
brightness profile for the extended models. The surface brightness
profile for the bluest ({\it dotted blue line}) and reddest ({\it red
solid line}) 15$\%$ of the photons are shown. The {\it left} and {\it
right panel} corresponds to models {\it 1.} and {\it 2.},
respectively. The figure shows that for collapsing gas clouds, the bluest Ly$\alpha$ photons may emerge more centrally concentrated than the reddest Ly$\alpha$ photons.}
\label{fig:bluevsred} 
\end{figure} 

Figure~\ref{fig:bluevsred} shows the surface brightness
profiles associated with model {\it 1.} ({\it left panel}) and
model {\it 2.} ({\it right panel}). The two profiles are
constructed from the reddest $15\%$ ({\it red solid line}) and the
bluest $15\%$ ({\it blue dotted line}) of all Ly$\alpha$ photons. The
bluest 15$\%$ is more concentrated, especially for model {\it 2.},
which has $\alpha=-0.5$. This suggests that steepening of the 
surface brightness profile towards bluer wavelengths within 
the Ly$\alpha$ line may be useful to diagnose gas infall. The amount
of steepening may constrain the infall velocity profile. 

\subsection{Summary of Intrinsic Ly$\alpha$ Properties}
\label{sec:summarylya}
The results obtained above can best be summarized as:

\begin{itemize}

\item {\it Result 1.} For all models, the Ly$\alpha$ radiation emerges predominantly blueshifted, typically by $500-3000$ km s$^{-1}$.This can be explained as follows: 
1) Photons propagating outwards in a collapsing gas cloud experience 
a net blueshift, because of energy transfer from the infalling gas
to the photons\footnote{According to equation~(\ref{eq:xinxout}), ${\vec v_a}\cdot {\vec k}_i$ is negative (${\vec v_a}$ and ${\vec k}_i$ are anti--parallel for photons that are propagating outwards), while ${\vec v_a}\cdot {\vec
k}_o$ is $0$ on average, which results in $x_0>x_i$.} 
2) Blueshifted
Ly$\alpha$ photons that are propagating outwards appear even bluer
(and thus further in the wing) for the infalling gas, which
facilitates their escape, while the reverse is true for redshifted
photons. These two effects yield a suppressed red and enhanced blue
peak. 

\item {\it Result 2.} The spectra of the central models are systematically bluer than those of the extended models. This result reflects that Ly$\alpha$ photons that are inserted deeper inside the collapsing gas cloud, obtain a larger average blue shift as they propagate outwards. That is, they have to traverse a larger column of neutral infalling gas, which allows more energy to be transferred from gas to photons, which results in a larger average blue shift. 

\item {\it Result 3.} The 'extended' models typically have an additional faint red peak in the spectrum separated by a minimum in the spectrum (the 'dip'). Both the location of the maximum of the red peak and the dip are uniquely determined by the infall velocity of the outermost gas layers. The reason for this is that some of the red Ly$\alpha$ photons that are propagating outwards are exactly at resonance in the frame of the infalling gas, which significantly reduces their escape probability and produces the observed dip. The dip in the spectrum is therefore strongly linked to the gas' infall velocity (at the outermost gas shells). In \S~\ref{sec:discussion} we point out that despite
the simplified nature of our models, in which the only bulk motions 
in the gas are radial and always in-flowing, this result is unlikely to change for more realistic, more complex gas distributions.

\item {\it Result 4.} The Ly$\alpha$ properties of the 'central' models
depend more sensitively on $\alpha$ than those of the 'extended' models, with
 the average blueshift of the spectrum strongly increasing 
with decreasing $\alpha$. This reflects that all Ly$\alpha$ photons in the central models are exposed to scattering by the gas in the central regions. Changing $\alpha$ from $1.0$ to $-0.5$, enhances $v_{\rm bulk}$ by a factor of $(r_{\rm vir}/r)^{1.5}$, which obviously becomes very large at small radii ($r \ll r_{\rm vir}$). Due to the enhanced infall velocity at small radii, more energy can be transferred from the gas to the photons in models with low values of $\alpha$, which results in a larger average blue shift of the Ly$\alpha$ line. 

\item {\it Result 5.} The prominence of the red peak, quantified by the ratio $[B]/[R]$, increases with decreasing infall velocity, mass and total Ly$\alpha$ optical depth. The ratio $[B]/[R]$ decreases with redshift for fixed other model parameters. These trends are easily understood: The larger the infall velocity, and the total Ly$\alpha$ optical depth, the more energy is transferred from gas to photons as the photons work their way outwards. This results in a more prominent blue peak and larger $[B]/[R]$. As explained above, increasing $\alpha$ reduces the number of very blue photons and, subsequently, $[B]/[R]$. The increase of $[B]/[R]$ with mass is caused by the increase of both the infall velocity, and the total Ly$\alpha$ optical depth with mass. The decrease of $[B]/[R]$ with redshift caused by the increased probability with redshift for photons on the blue side of the line center to be scattered out of our line of sight in the IGM. This is discussed in more detail in \S~\ref{sec:IGMI}.

\item {\it Result 6.} The surface brightness profiles for the extended models are typically flat, $\partial F /\partial {\rm ln} \theta=-1$ to $-0.5$, with an increasing flatness with increasing mass and $\alpha$.
The same trends are seen for the central models, which consistently
 have steeper surface brightness distributions with
$\partial F /\partial {\rm ln} \lsim -2$. One reason that the surface brightness profiles of the extended models become flatter with increasing $\alpha$ is that the intrinsic
emissivity becomes flatter with increasing $\alpha$ (see Fig~\ref{fig:panel1}).
This is not the only reason though, since the same trend is seen among the central models, which have identical $j(r)$ (apart from the amplitude). As was mentioned above, for lower values of $\alpha$ the number of very blue Ly$\alpha$ photons is enhanced, 
which increases the steepness of the surface brightness profile (see
 {\it result 7.} and its explanation below). 
The flattening of the surface brightness profile with mass is
caused by: 1) The projected emissivity on the sky becomes increasingly 
shallow with mass, which is due to the mass dependence of the virial radius
\footnote{If the projected emissivity were plotted as a function of
$\theta/\theta_{\rm vir}$, in which $\theta_{\rm vir}$ is the angle
the virial radius subtends on the sky, it would be
identical for all masses.} (for the extended models). 
2) The Ly$\alpha$ optical depth increases with
mass, which reduces the probability that photons propagate
through the outer gas layers without being scattered.

\item {\it Result 7.} A flattening of the surface brightness profile towards longer wavelengths within the Ly$\alpha$ line is indicative of infall. This can be explained as follows: the bluest photons emerge mainly from the center (as argued above) and are unlikely to get scattered in the outer gas layers.  This is because these photons appear even bluer in the frame of the infalling gas, and thus further in the line wing. The reverse is true for the reddest photons, which appear much closer to the line center in  the frame of the infalling gas and therefore likely scattered in the outermost gas layers. This produces a more diffuse appearance of the reddest photons compared to that of the bluest.

\end{itemize}

\begin{figure*}[t]
\vbox{ \centerline{\epsfig{file=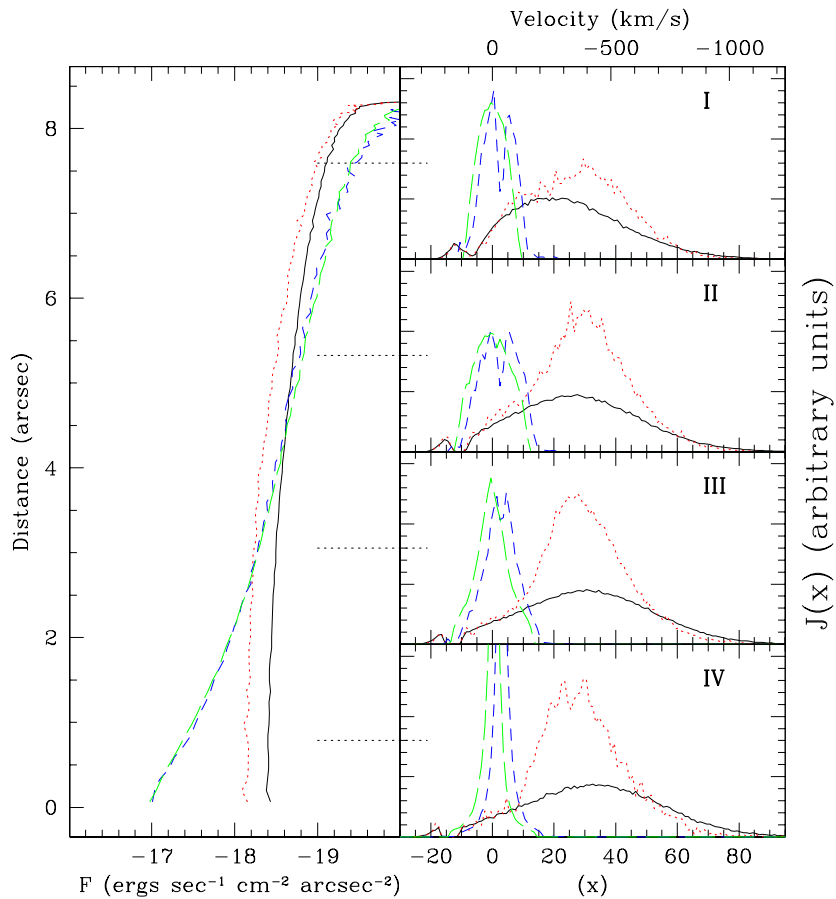,width=12.0truecm}}}
\caption[The effect of a central quasar on \lya characteristics.]{This
figure displays the \lya surface brightness distribution ({\it left panel})
and spectra emerging ({\it right panels}) from a collapsing gas cloud containing an ionizing source for four different ionizing luminosities. The spectra in the different panels are centered on the impact parameter indicated by the {\it black--dotted lines}. Model {\it 1.} is represented by the {\it
black--solid line} ($L_{\rm ion}=0$). The {\it red--dotted line}, {\it blue--dashed line} and {\it green--long--dashed line} corresponds to model {\it 1.} plus an additional central ionizing source with $L_{\rm ion}=5\times
10^{42}$ ergs s$^{-1}$, $L_{\rm ion}=10^{43}$ and $10^{46}$ ergs
s$^{-1}$, respectively. As $L_{\rm ion}$ increases, the spectral line becomes narrower and more symmetric around the line center.}
\label{fig:panelqso} 
\end{figure*}
\section{Fluorescent Ly$\alpha$ Emission from Self-Ionized Protogalaxies}
\label{sec:centralsource}

 In this section, we focus on the observable properties
of fluorescent Ly$\alpha$, such as its surface brightness distribution
and its spectrum. For this purpose, an ionizing point source was
inserted at the center of our sphere. Its total flux in ionizing
photons is $L_{\rm ion}$ (in ergs s$^{-1}$) and its spectrum is a
power law of the form $J(\nu,r)=J_0(r) (\nu/\nu_L)^{\beta}$.
 The slope $\beta$ depends on the nature
of the ionizing sources. We choose $\beta=1.7$, for the point source
to represent a quasar. The value of $J_0(r)$ is obtained by
$J_0(r)=(\beta-1)L_{\rm ion}/[16 \nu_L \pi^2 r^2]$. The standard
equations of photoionization equilibrium were solved numerically with
the code. The optical depth to the first shell was assumed to be $0$.
This was used to calculate its neutral fraction of hydrogen. This
allows the optical depth to the next shell to be computed, which also
determines its neutral fraction. This procedure is repeated until the
outer shell is reached. The resulting changes in the intrinsic 
Ly$\alpha$ spectrum are presented below.

\subsection{Impact of a Central Ionizing Source on the Properties of the Emerging Ly$\alpha$ Emission}
\label{sec:bh}

In the presence of an ionizing source we set the Ly$\alpha$ emissivity to be given by:

\begin{equation}
j(r)=\mathcal{F}_{\alpha}\alpha_B[1-x_H(r)]^2n_H^2(r)\hs 
h\nu_0+f_{\alpha}\frac{G \hs M_{\rm tot}(<r)\hs \rho_b}{r^2} \frac{dr}{dt}
\label{eq:jrqso}
\end{equation} 
where $\mathcal{F}_{\alpha}$ is the average number of Ly$\alpha$
photons produced per recombination, which is $\sim 68\%$ for case B and $\sim 42\%$ for case A recombinations. Case A/B applies when the gas is optically thin/thick to all Lyman series photons, respectively. 
The second term is the same as used for the
extended cooling models (eq.~\ref{eq:jr}). The main reason for
keeping this term is to make the transition from the previous
extended models to the current models smooth.

The gas in model {\it 1.} becomes fully ionized at $L_{\rm ion}
=10^{43}$ ergs s$^{-1}$. We caution that this estimate ignores
small--scale gas clumping. If we denote the clumping factor by $C
\equiv \langle \rho^2 \rangle/\langle \rho \rangle^2$, then this
transition to full ionization occurs at $\sim C$ times higher ionizing
luminosity (which would allow for $\sim C$ times higher Ly$\alpha$
luminosity) \footnote{ We mentioned in \S~\ref{sec:intro} that the total Ly$\alpha$ emissivity can be greatly boosted by embedded ionizing sources. A requirement for this boost is that the gas is clumpy, which increases the
recombination rate and allows a larger fraction of the ionizing
radiation to be reprocessed into Ly$\alpha$ photons, rather than
escaping, e.g. \citep[]{PP67,Haiman01,Alam02}. Because clumpiness is
not included in our model, this boost in the \lya luminosity is not
achieved.}.

Increasing $L_{\rm ion}$ beyond this value, merely
reduces the residual neutral fraction, and consequently, $N_{\rm HI}$.
The `phase transition' from neutral to fully ionized dramatically
affects the emerging Ly$\alpha$ spectrum. In Figure~\ref{fig:panelqso}
we show the Ly$\alpha$ spectrum and surface brightness distribution
for four values of $L_{\rm ion}$: $L_{\rm ion}=0$ ({\it black--solid
line}), $L_{\rm ion}=5\times 10^{42}$ ergs s$^{-1}$ ({\it red--dotted
line}), $L_{\rm ion}=10^{43}$ ergs s$^{-1}$ ({\it blue--dashed line})
and $L_{\rm ion}=10^{46}$ ergs s$^{-1}$ ({\it green--long--dashed
line}).  The results are presented differently than above: The {\it
left panel} shows the surface brightness distribution rotated by
90$^{\circ}$. The {\it right panels} show spectra of these models at
four different impact parameters, $\theta$, indicated by the {\it
horizontal dotted lines}. For example, the {\it lowest panel} shows
the spectrum emerging from $\theta \sim 0-2''$, the {\it second lowest
panel} shows the spectrum emerging from $\theta \sim 2-4''$, etc. The
main reason for presenting the data in this fashion is to study the
variations in the spectra across the object. Note that the spectra are
arbitrarily scaled for visualization purposes.

We note that in the the absence of an ionizing source, the spectrum
({\it black--solid line}) is fairly constant across the object, the
central region being slightly bluer. For low ionizing luminosities,
$L_{\rm ion}=5\times 10^{42}$ ergs s$^{-1}$, photo-ionization enhances
the Ly$\alpha$ emission from the central regions. Enhanced 
emission in the center enhances the number of photons in the 
bluest parts of the spectrum (\S~\ref{sec:summarylya}), and
steepens the surface brightness profile ({\it observation 6.} in
\S~\ref{sec:summarylya}). These are indeed the trends that can be seen
in Figure~\ref{fig:panelqso}. When
$L_{\rm ion}$ is increased by a further factor of two, the gas sphere as a
whole is fully ionized, which produces several effects. The
overall blueshift of the spectrum ({\it blue--dashed line })
almost completely vanishes, and its FWHM is significantly smaller.
 The main reason is the reduced column of neutral hydrogen, which reduces the
importance of broadening of the Ly$\alpha$ line by resonant
scattering. Instead, the broadening is determined by the bulk velocity
of the gas, despite the fact that each Ly$\alpha$ photon scatters
$\sim 10^4$ times before it escapes. The reason that the
Ly$\alpha$ spectrum is barely affected by scattering, is that each
Ly$\alpha$ photon mainly scatters near the site of its creation,
where the photon is at resonance. A chance encounter with a rapidly
moving atom can put the photon in the wing of the line, which allows
it to escape in a single flight from the local resonant region 
\citep[in which the optical depth is $\tau_0 \sim 10^3$-well within the regime of moderate opacities,][]{Adams72}. These photons escape
this region in a typical symmetric double peaked profile around the local line
center. Velocity gradients in the gas facilitate the photon's escape
from the rest of the object, effectively without 'seeing' shells at other radii. The frequency with which the photon
finally emerges at the virial radius therefore does contain
information on the bulk velocity of the site where the photon was
generated. The double peak in the emission line in regions
I-III is a consequence of
this local resonant scattering. The main reason for the enhanced
prominence of the dip at large impact parameters is that the
velocity gradient projected along the line of sight is smaller there,
which increases the total column of hydrogen a Ly$\alpha$ photon
effectively sees.

The reason that the spectrum from region IV is narrower is that 
this spectrum mainly contains photons that emerge from the central regions, in which the bulk infall velocity is low ($\alpha=1.0$). The effect of the
velocity field and $L_{\rm ion}$ ion the emerging spectrum is
investigated in more detail in \S~\ref{sec:bhonly}.

\subsection{Constraints from Ly$\alpha$ Emission on Gas Kinematics in the Presence of a QSO}
\label{sec:bhonly}

In the previous section we argued that the gas
kinematics can be constrained by the Ly$\alpha$ emission despite the
possibility that the Ly$\alpha$ photon scatters up to $\sim 10^4$ times. 
In this section we show that, although the amplitude of the bulk velocities
in the gas can be constrained well by the gas, the sign is much harder
to determine, since the emerging spectra for infall and outflow are
almost identical for fully ionized gas. In this section, we ignore
contribution from cooling to the emissivity; the second term on the
RHS in equation~(\ref{eq:jrqso}) is set to 0. The reason is that we
want to explore the similarities of the effects of scattering,
so we prefer to adopt the same emissivity profiles.

\begin{figure}[t]
\vbox{ \centerline{\epsfig{file=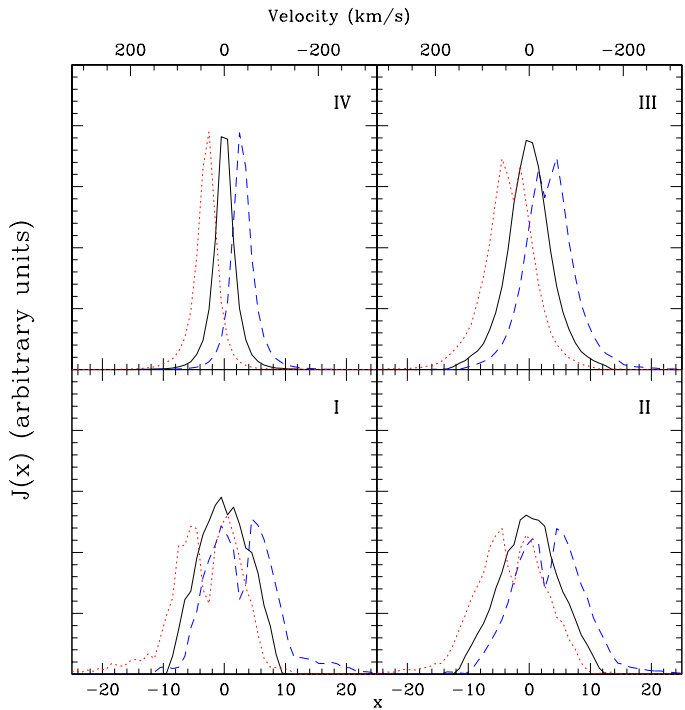,width=9.5truecm}}}
\caption[\lya spectra for optically thin infall vs. outflow.]{This figure shows the Ly$\alpha$ spectra emerging from optically thin outflowing and infalling gas. Each panel corresponds to the
impact parameter indicated by the roman letter (as in
Fig.~\ref{fig:panelqso}). The spectra shown by the {\it blue-dashed
lines} represent a photo-ionized infall model, while {\it red--dotted
lines} represent the outflow models in which the sign of $v_{\rm amp}$
is reversed. The {\it black--solid lines} represent both infall and
outflow models in the absence of scattering. The figure shows that when the gas is optically thin to Ly$\alpha$ photons, $\tau_0 \lsim 10^3$, infall and outflow models can produce nearly identical Ly$\alpha$ spectra, except for a small shift in frequency.}
\label{fig:panelqso2} 
\end{figure}

In Figure~\ref{fig:panelqso2} we show spectra at four different impact
parameters for three different models. The {\it blue--dashed lines}
represent model {\it 1.} containing a central ionizing source with
$L_{\rm ion}=10^{43}$ ergs s$^{-1}$ (as in Fig.~\ref{fig:panelqso}).
These spectra are compared to the no--scattering profiles, which refer
to the profiles that would be observed if the gas were transparent to
Ly$\alpha$ (shown as {\it black--solid lines}). The {\it red--dotted
lines} represent a model in which the sign of $v_{\rm amp}$ is
reversed. Each panel contains the spectra at the impact parameter
indicated by the roman numerals in the top--right corner (as in
Fig.~\ref{fig:panelqso}).
The only prominent difference between all spectra, is that the models
in which scattering occurs are shifted by several Doppler widths
relative to the no-scattering spectrum. Infall and outflow models are
slightly blue and redshifted shifted, respectively. The reasons for
this shift have been discussed in \S~\ref{sec:summarylya}.

Although not shown in this paper, we found the surface brightness
distributions for the optically thin infall and outflow models to be the identical (also see ZM02, who found the same). This leads to the conclusion that for gas that is only moderately thick to Ly$\alpha$ scattering, $\tau_0 \lsim 10^3$, the Ly$\alpha$ spectrum can be used to determine its velocity dispersion, despite the fact that Ly$\alpha$ photon up may scatter up to $10^3-10^4$ times. However, the {\it sign} of the velocity field can only be determined when the exact Ly$\alpha$ line center is known, to an accuracy of $\sim 10$ km s$^{-1}$ (which is probably not possible in
practice). This introduces an ambiguity in the interpretation of the Ly$\alpha$  blob found around a $z=3$ quasar by \citet{Weidinger04,Weidinger05}. Although they attribute the observed spectra and surface brightness profiles to gas infall, our results suggest that outflowing gas can produce similar characteristics. We will explore this issue in more depth in paper II.

\section{Discussion}
\label{sec:discuss}
\subsection{The IGM Transmission}
\label{sec:IGMI}
As mentioned in \S~\ref{sec:modelinput}, we have focussed on the intrinsic Ly$\alpha$ emission from collapsing protogalactic gas clouds. The subsequent
Ly$\alpha$ transfer through the IGM, which is defined to lie beyond the virial radius, has been ignored so far. Even though
the IGM is almost fully ionized at most redshifts considered in this paper, a
small residual neutral fraction is enough to scatter some fraction of
the Ly$\alpha$ photons from our line of sight.  For an IGM that is
comoving with the Hubble flow, the mean transmission is 1 and $\langle
e^{-\tau} \rangle$ redward and blueward of the line center,
respectively \citep[e.g.][]{Madau95}. Note that $\langle e^{-\tau}\rangle$ 
is to a good approximation constant, apart from the small-scale
  fluctuations in wavelength caused by the Ly$\alpha$ forest
 (i.e., overall redshift evolution is not significant).
 The exact value of $\langle e^{-\tau}
\rangle$ is a function of redshift and is well known from studies of
the Ly$\alpha$ forest \citep[e.g.][]{Songaila02,Lidz02,McDonald01} and
is tabulated in Table~\ref{table:IGM}.

\begin{table}[ht]
\small
\caption{Transmission of IGM}
\label{table:IGM}
\begin{center}
\begin{tabular}{lccccccc}
\tablewidth{1in}
z & 2.4 \tablenotemark{1} & 3.0 \tablenotemark{1}&3.8 \tablenotemark{1}& 4.5 
\tablenotemark{2}& 5.2 \tablenotemark{2}& 5.7 \tablenotemark{2}& 6.0\tablenotemark{2} \\
\hline
$f \equiv <e^{-\tau}>$ & 0.8 & 0.68 & 0.49 & 0.25 & 0.09 & 0.07 & 0.004 \\
$\sigma_f$ &   &  &  &  & 0.02 & 0.003 & 0.003 \\
\hline
\multicolumn{8}{l}{(1) From \citet{McDonald01}}\\
\multicolumn{8}{l}{(2) From \citet{Songaila02}}\\
\end{tabular}\\[12pt]
\end{center}
\end{table} 
The naive way to incorporate the impact of the IGM, is to give a photon
that escaped with $x>0$ and $x \leq 0$ a weight $\langle e^{-\tau}
\rangle$ and $1$, respectively. This is how the dependence of $[B]/[R]$
on redshift was calculated in \S~\ref{sec:redpeak} (see Fig.~\ref{fig:boverr}).

In \S~\ref{sec:lyaresults} we expressed the total Ly$\alpha$ luminosity,
$L_{{\rm Ly}\alpha}$, as a function of our model parameters. To correctly convert this to a flux on earth, denoted by $f_{{\rm Ly}\alpha}$, is not straightforward, due to the frequency dependence of the IGM transmission. Motivated by the observation that the \lya emerges mainly blueshifted from our models, we estimate the total detectable Ly$\alpha$ flux from:
$f_{{\rm Ly}\alpha}\approx <e^{-\tau}>L_{{\rm Ly}\alpha}/4 \pi
d_L^2$. We obtain:

\begin{eqnarray}
\frac{f_{{\rm Ly}\alpha}}{10^{-18}}\approx 2 \Big{(} 
\frac{M_{\rm tot}}{10^{11}M_{\odot}}\Big{)}^{\frac{5}{3}} 
\Big{(} \frac{v_{amp}}{v_{\rm circ}}\Big{)} 
\Big{(} \frac{2-\alpha}{1.75}\Big{)}^{1.2}\times \nonumber \\
\left\{ \begin{array}{ll}
         \ \times (\frac{5}{1+z_{\rm vir}})^{1.75}& z \leq 4.0\\
            \\
         \ \times (\frac{5}{1+z_{\rm vir}})^{8.3} & z > 4.0 
\end{array} \right.  
\hs \frac{{\rm ergs}}{{\rm sec}\hs{\rm cm}^2},
\label{eq:flux}
\end{eqnarray} 
where we used $<e^{-\tau}>=0.68[(1+z)/4]^{-1.5}$ for $z \leq 4$ and
$<e^{-\tau}>=0.031[(1+z)/7]^{-8.1}$ for $z > 4$, where the last
approximation was taken from \citep{Songaila02}. Note that this
approximation fails at $z \geq 6$, where the mean transmission drops
even more rapidly \footnote{To allow $f_{{\rm Ly}\alpha}$ to be written as a simple function of $(1+z)$, we approximated the luminosity
distance by $d_L(z)=$ $3.6 \times 10^{4}([1+z]/5)^{11/8}$ Mpc, which
is within $< 6\%$ of its actual value, in the range $z=3-10$.}. 
We caution that the strong break in the slope of the redshift dependence of the
 flux (eq.~\ref{eq:flux}) at $z > 4$ should not be taken literally.
In reality, the variation of the this slope with redshift should be smoother.
The main point of eq.~(\ref{eq:flux}) however, is that because the IGM becomes
increasingly opaque towards higher redshifts at $z \gtrsim 4$, cooling radiation of a fixed luminosity is more difficult to detect at $z \gtrsim 4$ than at $z \lsim 4$. Quantitatively, the total detectable Ly$\alpha$ cooling flux does {\it not} scale with $d_L^{-2}$ (which can be approximated as $\propto (1+z)^{-2.75}$), but as $ \propto <e^{-\tau}>d_L^{-2}$ (which can be approximated as $\propto(1+z)^{-11}$) for $z \geq 4.0$. We note that in the 
the proximity of bright QSO's, significantly
more flux may be transmitted \citep[e.g.][]{Cen00,Madau00,Cen05},

The above model of IGM transmission is
appropriate for an IGM that is comoving with the Hubble flow.
The gravitational pull exerted by the massive objects considered in this paper, causes their surrounding gas to decouple from the Hubble flow and move inwards. As a consequence, photons that escape our object with a small redshift, can Doppler shift into resonance in this infalling IGM, which increases their probability of being scattered out of the line of sight. The result of this is that the IGM may erase a fraction of the red part Ly$\alpha$ 
line \citep{Barkana042,Santos04}. We postpone
a detailed discussion of this effect to Paper II, but note that it
barely affects the quoted values of $[B]/[R]$ in Figure~\ref{fig:boverr}.

\subsection{Robustness of our Results}
\label{sec:discussion}

The simplified nature of our models clearly introduces uncertainties
to the results presented in this paper. Nevertheless, several results
obtained in this paper should be robust. The conclusion that \lya
emerges blue--shifted from a collapsing cloud will not change for
different gas distributions, since it simply relies on \lya photons
encountering gas with a bulk velocity that opposes their propagation
direction. The only requirement is that the Ly$\alpha$ photons do meet
gas on their way out. If the gas is very clumpy, and only covers part
of the sky from the site where the Ly$\alpha$ is generated, then it
could escape without a net blue shift. For the massive protogalactic
halos considered in this paper the covering factor is expected to be
well above one, and we do not expect Ly$\alpha$ to escape without
encountering any gas. As argued by \citet{Birnboim03}, virial shocks
may be absent in halos with $M_{\rm crit} \lsim 2 \times 10^{11}
M_{\odot}$, which suggests that even if the gas is clumpy, in the
protogalactic phase, the clumps are not embedded in a hot tenuous
medium for masses below $M_{\rm crit}$. Although the mass of our
fiducial model lies slightly above this quoted critical mass, its
exact value depends on redshift, metallicity of the gas, etc. In this
case the gas is also optically thick to Ly$\alpha$ photons between the
denser clumps, making our calculations more appropriate.

The presence of the dip in the emergent spectrum redward of the line
center ({\it Result 3.} in \S~\ref{sec:summarylya}) is most likely not
an artifact of the simplified nature of our models.  The
representation of the gas dynamics in our models, as a smooth and
radial infall, is a clear oversimplification.  In reality, gas infall
is likely to be clumpy, and may have significant tangential velocity
components. However, even for this type of complex gas distribution
and velocity field, a the dip is likely to be present, since it only
requires Ly$\alpha$ photons redward of the line center to be at
resonance in the frame of the gas in the outermost shells. Since the
location of the dip will depend on the local velocity of the gas in
the outermost shells, the position of the dip would vary across the
object for more complex gas motions. However, as long as the gas cloud
as a whole is collapsing, the overall escape probability for
Ly$\alpha$ photons redward of the line center is reduced, which may
show up as a noticable (but broader) dip. Observational support for
this claim is that \citet{Barkana03} have found evidence for the
presence of a dip in the Ly$\alpha$ spectra of several quasars, which
is believed to be caused by infalling gas.

Similarly, that the bluest Ly$\alpha $photons can freely stream outwards through a column of collapsing gas, while the reddest Ly$\alpha$ photons get scattered in the outermost regions, should be valid regardless of the exact gas
distribution. It is not clear however, what fraction of centrally
produced Ly$\alpha$ escapes. The presence of even a small amount of
dust is sufficient to absorb Ly$\alpha$ photons 
\citep[][]{Neufeld90,Charlot91}.
 If the medium is clumpy, then Ly$\alpha$ photons may
still escape by scattering of (dusty) cold clouds embedded in a
tenuous medium \citep[][]{Neufeld91,HaimanSpaans99,Hansen06}.
 The red peak, which is produced further out, suffers less from
dust absorption. The net effect of destroying preferentially the bluer
photons would weaken the spectral dependence of the surface brightness
distribution for the extended models.
Therefore, although a flattening of the surface brightness distribution with increasing wavelength (within the Ly$\alpha$ line) is indicative of gas infall, its absence, however, does not argue against infall. Once detected, the amount of flattening could constrain the infall velocity profile.
Detailed results, such as the exact shapes of the spectra,
 are very likely to change for realistic gas configurations.
For reliable detailed predictions, we should apply the code to the output of
cosmological simulations which can provide the Ly$\alpha$ code with
realistic density, velocity and temperature fields \citep[as in][]{Iro05}. However, the calculations presented in this paper relating to peak morphology do offer simple insights in what may be plausible Ly$\alpha$ properties of collapsing protogalaxies. In PaperII, we demonstrate the
  usefulness of our models, by applying them to several observations.

The gas temperature in this paper was assumed to be $10^4$ $^{\circ}$K.
Turbulent motions in the gas may raise the effective temperature of the gas, which was the reason ZM02 used $T\sim 10^5$ $^{\circ}$K. The exact temperature of the gas affects the frequency redistribution
of Ly$\alpha$ photons undergoing resonant scattering and
therefore affects the results in which the broadening and blueshift of
the Ly$\alpha$ emission line are dominated by resonant
scattering. However, the temperature dependence of our results is
weak. This is best illustrated for the case of a static,
uniform sphere, in which the emergent spectrum is entirely determined by resonant scattering. If all Ly$\alpha$ photons are injected in the center of
the sphere at $x=0$, the emergent spectrum peaks
 at $x_p= \pm 0.92 (a \tau_0)^{1/3}$
(\S~\ref{basics}). For a fixed total column of hydrogen from the
center to the edge of the sphere, the velocity offset from the line center 
of the two peaks depends on temperature as: $v_{\rm shift} \propto
T^{1/6}$. A change in temperature by a factor of 10 shifts our lines
by a factor of $\lsim 1.5$.

\subsection{Comparison with Previous Work}
\label{sec:otherwork}

In this paper we used a Monte Carlo technique that is similar
 to those used in previous papers \citep[e.g.][]{Ahn00,Ahn02,Zheng02,Cantalupo05}. The following discussion briefly summarizes how our work differs from and
complements the work previously done. \citet{Ahn02} and \citet{Ahn03} have focussed primarly on static, extremely thick slabs which represented star burst galaxies and optically thick expanding superwinds. The Ly$\alpha$ from their superwind models emerged with a net redshift, for the same reason our models predict blueshifted Ly$\alpha$ emission. \citet{Zheng02} concentrated on fluorescent Ly$\alpha$ emission from Damped Lyman $\alpha$ (DLAs) systems illuminated by an externally generated UV background (either diffuse, or produced by a nearby quasar). They found self--shielding to produce a core in the surface brightness profiles, which differs from the surface brightness profiles of any of our models. Additionally, the spectra emerging from their fluorescent DLAs are double peaked and symmetric around the line center. However, this symmetry is mainly due to the fact that \citet{Zheng02} choose their gas cloud to be static. \citet{Cantalupo05} also considered fluorescent Ly$\alpha$ radiation from high column density systems using high resolution hydro-simulations. They found the symmetry in the spectra of fluorescent Ly$\alpha$ radiation to vanish for more realistic velocity fields, e.g. collapsing gas clouds produce a more prominent blue peak, similar to the spectra presented in this paper. However, fluoresent emission in response to the diffuse ionizing background only reaches flux levels of $\sim 10^{-20}$ ergs s$^{-1}$ cm$^{-2}$, which is well below most of the signals calculated in this paper. The amount of fluorescent Ly$\alpha$ emission can be boosted significantly in the presence of a nearby bright quasar, which should be easily identifiable. It appears possible to observationally distinguish between our models and those discussed above.


\section{Summary and Conclusions}
\label{sec:conclusions}

We presented Ly$\alpha$ Monte Carlo radiative transfer calculations
through neutral (\S~\ref{sec:cooling}), and self-ionized
(\S~\ref{sec:centralsource}), collapsing, spherically symmetric gas
clouds. To provide a test case for our code that is tailored to
spherical symmetry, we derived a new analytic solution for the
Ly$\alpha$ spectrum emerging from a uniform, static sphere
(\S~\ref{sec:test}, Appendix~\ref{sec:analytic}). We found that a
cosmological abundance of deuterium regulates the transfer of
Ly$\alpha$ photons over a narrow frequency range (corresponding to
$40-50$ km s$^{-1}$ at $T=10^4$ $^{\circ}$K, see \S~\ref{Deut}). Although the
results in the present paper are not affected by the presence of
deuterium, we point out it may affect the analysis for systems with
hydrogen column densities in the range $10^{18}-10^{20}~{\rm
cm^{-2}}$, for $T=10^4$ $^{\circ}$K.

We studied the effects of gas kinematics and Ly$\alpha$ emissivity 
as a function of radius, $j(r)$, on the Ly$\alpha$ spectrum emerging from
the collapsing clouds and on the Ly$\alpha$ surface brightness distribution,
for a range of models with different masses and redshifts
(\S~\ref{sec:closerlook}), which are summarized in Table~\ref{table:models}

 The emergent Ly$\alpha$ spectrum is typically double--peaked and asymmetric. In practice, we find that the blue peak is significantly enhanced, which results in an effective blueshift of the Ly$\alpha$ line. This blueshift may easily be as large as $\sim 2000$ km s$^{-1}$ (\S~\ref{sec:bluepeak}). The total blueshift of the line increases with the total column density of neutral hydrogen of the gas and with its infall speed (\S~\ref{sec:bluepeak}, Fig~\ref{fig:fwhm}). The prominence of the red peak is enhanced towards higher redshift, lower mass and lower infall speeds (\ref{sec:redpeak}, Fig~\ref{fig:boverr}). If detected, the shift of the red peak relative to the line center may give a potentially useful measure of the gas infall speed (\S~\ref{sec:redpeak}, Fig~\ref{fig:redpeak}). Furthermore, a steepening of the surface brightness profile towards bluer wavelengths within the Ly$\alpha$ line is a useful diagnostic for gas infall. The amount of steepening may constrain the infall velocity profile (\S~\ref{sec:surfacebrightness}, Fig~\ref{fig:bluevsred}). A more complete summary of all results involving the surface brightness profiles, the morphology of the blue and red peak, and their relative prominence is given-together with a qualitatitive explanation-in \S~\ref{sec:summarylya}.

The `fluorescent' (or 'self-ionized') cases (\S~\ref{sec:centralsource}) relate to models of collapsing gas clouds with an embedded ionizing source. As its ionizing luminosity increases, the Ly$\alpha$ line becomes narrower and more symmetric around the line center (Fig~\ref{fig:panelqso}). When the gas is fully ionized, we found the line widths in the spectrum to be a good measure for the velocity distribution of the gas, despite the fact that in some cases each Ly$\alpha$ photon may be scattered up to $10^3-10^4$ times. We caution that for these optically thin cases, infall models have almost the same spectrum as outflow models (and identical surface brightness profiles, also see ZM02). The only difference is that for infall/outflow the emission line is shifted in frequency to the blue/red, respectively (Fig.~\ref{fig:panelqso2}), by an amount which
decreases with the brightness of the central ionizing source. We will explore this issue in more depth in paper II.

The effective blueshift of the Ly$\alpha$ emission from 
collapsing protogalaxies has profound consequences for its
detectability, as radiation blueward of the Ly$\alpha$ line is subject
to IGM absorption (\S~\ref{sec:IGMI}), an effect that increases rapidly with redshift ($e^{-\tau} \propto (1+z)^{-8}$ for $z\geq 4$). We expect
the Ly$\alpha$ fluxes from high redshift ($z \gtrsim 4.0$) collapsing
protogalaxies to be very weak, unless the intrinsic Ly$\alpha$
luminosity is very high (eq.~\ref{eq:flux}). In Paper II, where we use
the main results of this paper to interprete observations of Ly$\alpha$ emitters, the impact of the IGM on our results is discussed in more detail.

Ongoing and future surveys of Ly$\alpha$ emitters are providing us
with useful constraints on the early stages of galaxy formation. We
hope the work presented in this paper can contribute to the
interpretation of recent and of forthcoming observations of Ly$\alpha$
emitting galaxies.\\

{\bf Acknowledgements} We thank Mike Shara for suggesting to consider deuterium, Jim Applegate for numerous insightful comments; Zheng Zheng for useful comments on an earlier version of the manuscript. Furthermore, we thank Adam Lidz, Martin Rees, and Rashid Sunyaev for useful discussions. MD
thanks the Kapteyn Astronomical Institute, and ZH thanks the
E\"otv\"os University in Budapest, where part of this work was done,
for their hospitality. ZH gratefully acknowledges partial support by
the National Science Foundation through grants AST-0307291 and
AST-0307200, by NASA through grants NNG04GI88G and NNG05GF14G, and by
the Hungarian Ministry of Education through a Gy\"orgy B\'ek\'esy
Fellowship. We thank the anonymous referee for his or her comments that improved the presentation of both our papers.


\begin{appendix}
\section{The Models}
\label{app:table}
The table below contains all models that we studied in this paper.
For each model, its associated set of model
parameters $M_{\rm tot}$, $z_{\rm vir}$, $\alpha$ and $v_{\rm amp}$ is
given. The infall velocity, $v_{\rm amp}$ has units of 
the circular velocity for (given in the fourth column). The first 
column contains the model number. Two models that only differ in their
choice of $j(r)$ have the same model number, but the  
 `central' (see the discussion above) model has an additional label 'c' attached . For reference, the last column contains the figure numbers in which results pertinent to each model are shown.

\begin{table}[ht]
\small
\caption{Models Studied}
\label{table:models}
\begin{center}
\begin{tabular}{ccccccc}
\tablewidth{3in}
 Mod. \# & $M_{\rm tot}$  & $z_{\rm vir}$ & $v_{\rm circ}$ &$v_{\rm amp} $ & $\alpha$ & Fig \#.\\
Cont. & $10^{11}M_{\odot}$  & & km s$^{-1}$ & $v_{\rm circ}$  &  \\
\hline
\hline
1  & 5.2 &3.0 & 181 & 1.0       &  1.0  & \ref{fig:panel1}--\ref{fig:surbrightness}\\
2  & 5.2 &3.0 & 181 & 1.0       & -0.5  & \ref{fig:panel1},\ref{fig:fwhm},\ref{fig:boverr},\ref{fig:surbrightness}\\
3 & 5.2 &3.8 & 198 & 1.0       &  1.0   &\ref{fig:boverr} \\
4 & 5.2 &4.5 & 212 & 1.0       &  1.0   &\ref{fig:boverr}\\
5 & 5.2 &5.2 & 225 & 1.0       &  1.0  & \ref{fig:boverr},\ref{fig:redpeak}\\
6  & 5.2 &6.0 & 240 & 1.0       &  1.0 & \ref{fig:boverr}\\
7 & 0.1 &3.0 & 49  & 1.0       & -0.5 &  \ref{fig:redpeak},\ref{fig:surbrightness}\\
8 & 0.1 &3.0 & 49  & 1.0       &  0.25 & \ref{fig:surbrightness} \\
9 & 0.1 &3.0 & 49  & 1.0       &  1.0  & \ref{fig:boverr},\ref{fig:surbrightness}\\
10 & 40  &3.0 & 355 & 1.0       &  1.0  & \ref{fig:boverr},\ref{fig:redpeak},\ref{fig:surbrightness}\\
11 & 40  &3.0 & 355 & 1.0       &  0.25 & \ref{fig:surbrightness}\\
12 & 40  &3.0 & 355 & 1.0       & -0.5  & \ref{fig:surbrightness}\\
13 & 5.2 &3.0 & 181 & 1.0       &  0.25 & \ref{fig:fwhm},\ref{fig:boverr},\ref{fig:surbrightness}\\
14 & 5.2 &3.0 & 181 & 0.50      &  1.0  & \ref{fig:fwhm},\ref{fig:boverr},\ref{fig:redpeak}\\
15 & 5.2 &3.0 & 181 & 0.25      &  1.0  & \ref{fig:fwhm},\ref{fig:boverr},\ref{fig:sequence}\\
16 & 5.2 &3.0 & 181 & $10^{-6}$ &  1.0  & \ref{fig:sequence}\\
17 & 5.2 &3.0 & 181 & 2.0       &  1.0  & \ref{fig:fwhm},\ref{fig:boverr}\\
18 & 5.2 &3.0 & 181 & 1.3       &  1.0  & \ref{fig:redpeak} \\
\hline
\hline
\end{tabular}\\[12pt]
\end{center}
\end{table}

\onecolumn

\section{Tests of the Code}
\label{app:test}

\subsection{Testing Core Scattering}
\label{app:testcore}
The test discussed in \S~\ref{sec:test} mainly addresses scattering of
photons in the wing, since $> 99\%$ and $>99.98 \%$ of the photons
that emerge from the optically thick sphere are in the wings, for
$\tau_0=10^5$ and $10^7$, respectively. To calculate these numbers,
the transition from core to wing was defined at $|x|=3$. Therefore,
the emerging spectrum is insensitive to the performance of the code
for scatterings in the core. Note that this is exactly why it is
allowed to deploy the accelerated scheme (\S~\ref{sec:fastcode}) to
calculate emerging Ly$\alpha$ spectra from optically thick media. To
test the accuracy of the code for scatterings in the core we compare
the redistribution function, $q_{II}(x,x')$ (\S~\ref{basics}), as
extracted from the code, with the theoretical solutions given by
\citet{Hummer62} and \citet{Lee74} (their $q_{IIb}(x,x')$).  In the
{\it upper right panel} of Figure~\ref{fig:test} we show the cases
with $x=0$, $x=2$ and $x=5$. The agreement between the histograms
generated by the code, and the theoretical curves is excellent.

An additional test for core scatterings involves the well studied
problem of the mean number of scatterings a Ly$\alpha$ photon
undergoes, $\bar{N}$, before it escapes from the slab
\citep[see][]{Adams72}. To calculate the average number of scatterings
a \lya photon undergoes in our code, we use the weighting scheme
described by \citet{Avery68}. This weighing procedure reduces the
average number of scatterings by a factor of $\sim 2$. The main reason
is that the distribution of the number of scatterings has a long tail
to large number of scatterings, which increases $\bar{N}$. The
weighting scheme described by \citet{Avery68}, gives less weight to
the photons that are in the high-$N$ tail, which brings the number of
scatterings of individual photons closer to the mean. The  dots
in the {\it lower left panel} denote $\bar{N}$ as a function of
$\tau_0$ as extracted from our simulation, whereas the solid line is
the theoretical prediction by \citet{Harrington73} $\bar{N}=1.61
\tau_0$. The agreement at large $\tau_0$ is impressive, but breaks
down when $\tau_0 \lsim 10^5$, which is in the transition regime
between moderate and extreme optical depths.\\

\subsection{Testing the Code when the Gas is not Static}
\label{sec:RL}
To test the validity of the code in a case when the gas has bulk
motions, we compare our results with LR99, who used a Monte Carlo
method to calculate the Ly$\alpha$ transfer through a fully neutral,
homogeneous, uniform intergalactic medium that undergoes Hubble
expansion. They demonstrated that ionizing sources prior to
reionization, are embedded in large Ly$\alpha$ halos that are
redshifted relative to the host galaxy. Initially, resonant scattering
moves the Ly$\alpha$ photons into both the red and blue wing of the
line profile, which dramatically increases their mean free path (as
discussed in \S~\ref{basics}). The increased mean free path enables
Hubble expansion to redshift the Ly$\alpha$ photons during their
flight. The blue photons return to the line core, while the red
photons become even redder. The Ly$\alpha$ photons that have
redshifted far enough from the line center stream freely to the
observer. LR99 use a simple Monte Carlo approach to calculate the
spectrum (their Fig. 2) and surface brightness profiles (their Fig. 1)
of these Ly$\alpha$ halos. We quantify this below:

A photon initially at frequency $x$ (in this case $x\ll 0$),
propagating through an expanding IGM, redshifts in proportion to its
traversed distance, $dx=-H(z)ds/v_{\rm th}$. Here, $H(z)$ is the
Hubble parameter at redshift $z$, $ds$ is the differential traversed
physical distance, and $v_{\rm th}$ is the thermal velocity of the
atoms in the IGM. As the photon redshifts, its mean free path
increases, and consequently Hubble expansion redshifts the photon even
further, etc. The maximum optical depth a photon initially at
frequency $x$ can see through a uniform IGM is given by:

\begin{equation}
\tau_x=\int_0^{\infty}ds\hs n_H(s)\sigma(s)=n_H
\sigma_0\int_0^{\infty}ds\phi(x'[s])= \frac{n_H
\sigma_0}{\sqrt{\pi}}\int_0^{\infty}ds\frac{a}{x'^2[s]}=\frac{-n_H
\sigma_0 v_{\rm
th}a}{H(z)\sqrt{\pi}}\int_x^{-\infty}\frac{dx'}{x'^2}=\frac{-n_H\sigma_0
a v_{\rm th}}{\sqrt{\pi}H(z) x}
\end{equation} 
For $T=10$ \kel, $z=10$, and the \wmap cosmological parameters,
\begin{equation}
\tau_x \approx \frac{-3.8 \times 10^3}{x}\equiv \frac{x_*}{x},
\end{equation} 
in which $x_*$ is the `critical' frequency, which marks the frequency
beyond which the neutral IGM is optically thin. Photons with $x \ll
x_*$ (note that $x$ is negative ) stream freely through the neutral
IGM towards the observer. LR99 define their frequency variable in
terms of this critical frequency as: $\hat{x}\equiv x/x_*$. Note that
blue- and redshifted photons have a negative and positive $\hat{x}$,
respectively.

We caution that the results presented in LR99 are not exactly adequate
to test our code in the case of bulk motions. In LR99 the Ly$\alpha$
photons redshift during their flight, while the energy before and
after scattering is equal (i.e.\ $x_o=x_i$). In the case the shells
have bulk motion, the photon's energy changes in a scattering event
according to (eq.~\ref{eq:xinxout}), while the photon's energy remains
constant during its flight. For example: A photon is emitted at $x=0$
and $r=0$ in the direction ${\vec k}$. Consider the situation in which
each scattering leaves ${\vec k}$ unchanged. In LR99, the photon
continuously redshifts in between fully coherent scatterings. In the
situation in which the gas shells are radially expanding (according to
Hubble's law), the photons frequency will remain the same, $x=0$. Note
that the mean free path of this photon is the same in both
situations. This photon emerges at a hypothetical boundary in the same
number of scatterings in both cases, but in LR99 it is redder. The
situation sketched above is very rare, but it is easy to convince
oneself that {\it any} photon, along an arbitrary trajectory, emerges
slightly redder in LR99's scenario. In practice, this difference is
small, but leaves a noticeable difference in the spectrum. 

LR99 clearly describe their Monte Carlo simulation, which can easily
be reproduced and adequately modified to describe Ly$\alpha$ transfer
through gas radially expanding outwards from $r=0$ according to the
Hubble law, $v_{\rm bulk}=H \hs r$, where $H$ is the Hubble
constant. The calculated spectrum for this modified LR99 simulation is
slightly bluer than their original spectrum shown in {\it their}
Figure 2. This modified curve is shown in the {\it lower right panel}
of {\it our} Figure \ref{fig:test} as the {\it green--solid}
line. Note that LR99 used dimensionless units, with all cosmological
parameters scaled out. The histogram shown was computed with our own
code. Ly$\alpha$ photons were injected in the center of a sphere in
which the shells are expanding outwards according to Hubble's law,
$v_{\rm bulk}=H(z) \hs r$. In our code, the variables are not
dimensionless. We pick $z_s=10$ (which is the physical model LR99
translate their results to) and the density of the gas shells is
uniform and $n_H \sim 2 \times 10^{-7}$ $(1+z_s)^3$ cm$^{-3}$. The gas
temperature was set to $10$ K ($a=1.5 \times 10^{-2}$, although the
exact temperature barely matters). To facilitate comparison with LR99n,
this plot uses their frequency variable, $\hat{x}\sim -3.8 \times 10^3
\hs x$. As the figure shows, the agreement is excellent.

\section{Analytic Solution for a Static Uniform Sphere}
\label{sec:analytic}

In this Appendix we derive an analytic solution for the Ly$\alpha$
spectrum emerging from a uniform, static sphere of gas, similar to the
well known solution for a plane--parallel slab derived by
\citet{Harrington73} and \citet{Neufeld90}.  Following
\citet{Harrington73,Neufeld90}, our starting point is the radiative
transfer equation after having made the Eddington approximation. In
spherical coordinates this reads:

\begin{eqnarray}
\frac{\partial^2 J(r,x)}{\partial r^2}+\frac{2}{r}
\frac{\partial J(r,x)}{\partial r}=
3 \phi^2(x)\kappa^2_0 (J(r,x)-3\phi(x)\kappa_0^2L(r), \nonumber \\
L(r)=\int_{-\infty}^{\infty}
\phi(x')q(x,x')J(x',r)dx'+\phi(x)\frac{j(r)}{4\pi}.
\end{eqnarray} 
\citet{Rybicki94} showed that the integral operator can be written
as a differential operator, after applying the {\it Fokker-Planck
approximation}:

\begin{equation}
\int_{-\infty}^{\infty}\phi(x')q(x,x')J(x',r)dx' \approx
\phi(x)J(x,r)+\frac{1}{2}\frac{\partial}{\partial x}
\phi(x)\frac{\partial J(x,r)}{\partial x}.
\end{equation} 
After defining the variable $\sigma$, by $\partial x/\partial
\sigma=(3/2)^{1/2}\phi(x)$ (note that in the line wing,
$\phi(x)=a/[\pi^{1/2}x^2]$, and therefore
$\sigma=(2\pi/27)^{1/2}x^3/a$), the transfer equation simplifies to:

\begin{equation}
\frac{\partial^2 J(r,x)}{\partial r^2}+\frac{2}{r}\frac{\partial
J(r,x)}{\partial r}+ \kappa_0^2 \frac{\partial^2 J(r,x)}{\partial
\sigma^2}=-3\phi^2(x)\kappa_0^2 \frac{j(r)}{4\pi}.
\label{eq:radtrans}
\end{equation} 
Following \citet{Unno55}, this equation is solved using the {\it
Sturm-Liouville} expansion which consists of an eigenfunction
expansion of the form:

\begin{equation}
J(r,x)=\sum_{n=1}^{\infty} E_n(r)h_n(x).
\end{equation} 
Let each eigenfunction, $E_n(r)$, obey the homogeneous differential
equation:

\begin{equation}
\frac{\partial^2 E_n(r)}{\partial r^2}+\frac{2}{r}
\frac{\partial E_n(r)}{\partial r}+\lambda^2 E_n(r)=
0 \hs \rightarrow \hs E_n(r)=C
\frac{\sin \lambda r}{\lambda r}+D\frac{\cos \lambda r}{\lambda r},
\label{eq:lambda1}
\end{equation} 
where $C$ and $D$ are arbitrary constants. Note that to prevent the
solution from diverging at $r=0$, $D=0$. The value of $\lambda$ is
obtained from the boundary condition. The flux can be written as
\citep[e.g.][ although the first authors have an extra factor of $\pi$ in the definition of the flux $F$]{RL79,Unno55,Harrington73,Neufeld90}:

\begin{equation}
F(r,x)=-\frac{-4}{3\kappa_0\phi(x)}\frac{\partial J(x,r)}{\partial r}
\label{eq:bound}.
\end{equation} 
At the boundary, where the radius is denoted by $r=R$, all photons are
propagating outwards, i.e.\ no photons get scattered back through the
surface. In this case, the flux $F(R,x)=2J(R,x)$. For this equality
 $b(r)=0$ was assumed. Each function $E_n(r)$ should
obey the boundary conditions, which gives the following constraint on
$\lambda$:

\begin{equation}
\lambda \cot \lambda R =-\frac{3}{2} \phi(x)+\frac{1}{R} 
 \rightarrow  \lambda_n  = \frac{1}{R}\pi n\Bigg{(}1- \frac{2}{3\phi(x)\kappa_0 R+2}+O[1/(3\phi(x)\kappa_0 R+2)^2] \Bigg{)},\hspace{5mm}  n =  1,2,3,...
\label{eq:lambda2}
\end{equation}  
The sequence of solutions for $\lambda$ was obtained after making the
same approximations as in \citet{Unno55,Harrington73}.  Solutions with
different values of $n$ are orthogonal. Each $E_n(r)$ can be
normalized by multiplying the function by $\lambda_n/\sqrt{2\pi
R}$. Then $J(r,x)$ is:

\begin{equation}
J(r,x)=\sum_{n=1}^{\infty} C_n\hs h_n(x)\hs 
\frac{\sin \lambda_n r}{\lambda_n r}=
\frac{1}{\sqrt{2\pi R}}\sum_{n=1}^{\infty}
\frac{{\rm sin} (\lambda_n r)}{ r}h_n(\sigma).
\label{eq:jrx}
\end{equation} 
We substitute this back into the radiative transfer equation
(eq.~\ref{eq:radtrans}). Following \citet{Harrington73}, the function
$3\phi^2(x)$ (which is peaked strongly around $x=0$) is replaced by
$\sqrt{6}\delta(\sigma)$. The solution for any particular
$h_m(\sigma)$ can be be found by multiplying the radiative transfer
equation on both sides by the $m^{\rm th}$ eigenfunction,
$\sin(\lambda_m r)/r\sqrt{2\pi R}$ followed by integration over $\int
4\pi r^2dr$:

\begin{eqnarray}
\kappa_0^2\frac{\partial^2}{\partial
\sigma^2}h_m(\sigma)-\lambda_m^2h_m(\sigma)=
\frac{-\sqrt{6}\kappa_0^2}{\sqrt{2\pi
R}}\frac{\delta(\sigma)}{4\pi}Q_m, \nonumber \\ 
Q_m=\int_0^R 4\pi r^2
dr \frac{{\rm sin} (\lambda_m r)}{r}j(r).
\end{eqnarray} 
The solution for $h_m(\sigma)$ is given by:

\begin{equation}
h_m(\sigma)=\frac{\sqrt{6}\kappa_0}{8\pi\sqrt{2\pi R}}Q_m\frac{{\rm
exp}(-\lambda_m|\sigma|/\kappa_0)}{\lambda_m},
\end{equation} 
which in combination with equation~(\ref{eq:jrx}) completes the
solution for $J(r,x)$.

\subsection{Analytic Solution for a Central Source.}

First consider the situation in which all photons are emitted at
radius $r_s$, then:

\begin{equation}
j(r)=\delta(r-r_s),\hspace{5mm} Q_m=\frac{{\rm sin} (\lambda_m
r_s)}{r_s},
\end{equation} 
when substituted into equation~(\ref{eq:jrx}) we get:

\begin{eqnarray}
J(r,\sigma)
&=&\frac{1}{\sqrt{2\pi R}}\sum_{n=1}^{\infty}\frac{{\rm sin}
(\lambda_n r)}{ r}\Big{(}\frac{\sqrt{6}\kappa_0}{8\pi \sqrt{2\pi
R}}Q_m\frac{{\rm
exp}(-\lambda_n|\sigma|/\kappa_0)}{\lambda_n}\Big{)}\nonumber\\
&=& 
\frac{\sqrt{6}}{16 \pi^2 R}\frac{\kappa_0}{r
r_s}\sum_{n=1}^{\infty}{\rm sin} (\lambda_n r){\rm sin} (\lambda_n
r_s)\frac{{\rm exp}(-\lambda_n|\sigma|/\kappa_0)}{\lambda_n}.
\end{eqnarray} 
The spectrum at the boundary is obtained by setting $r=R$. From
equation~(\ref{eq:lambda2}) we know:
\begin{equation}
{\rm sin} (\lambda_n R)={\rm cos} (\lambda_n R)\frac{\lambda_n
R}{1-1.5\phi(x)R\kappa_0}\sim (-1)^n
\frac{\lambda_n}{1.5\phi(x)\kappa_0}.
\end{equation} 
Therefore,

\begin{equation}
J(R,\sigma)=\frac{\sqrt{6}}{16 \pi^2 R^2}\sum_{n=1}^{\infty} \frac{2
\sqrt{\pi} x^2}{3a\kappa_0}(-1)^n {\rm
exp}\Big{(}\frac{-n\pi}{R}\frac{|\sigma|}{\kappa_0}\Big{)}\frac{{\rm
sin}\lambda_nr_s }{r_s}.\\
\end{equation} 
Using 
\begin{equation}
\sum_{n=1}^{\infty}(-1)^n x^n=\frac{-x}{1+x}, \hspace{2mm}{\rm for}
\hs |x|< 1 \hspace{2mm}, 
 \cos x  = \frac{e^{ix}+e^{-ix}}{2},\hspace{2mm} \sin
x=\frac{e^{ix}-e^{-ix}}{2i}, \hspace{2mm} \cosh
x=\frac{e^{x}+e^{-x}}{2},
\end{equation} 
we find:

\begin{eqnarray}
J(R,\sigma)=\frac{\sqrt{6}}{32 i \pi^2 R^2 r_s}\frac{2 \sqrt{\pi}
x^2}{3a\kappa_0} \Big{(}\frac{1}{1+{\rm
exp}(\frac{\pi|\sigma|}{R\kappa_0}+i\pi r_s/R)}- \frac{1}{1+{\rm
exp}(\frac{\pi|\sigma|}{R \kappa_0}-i\pi r_s/R)}\Big{)}\nonumber =\\
\frac{\sqrt{6}}{48 \pi^{3/2} R^2 r_s}\frac{x^2}{a\kappa_0}
\Big{(} \frac{{\rm sin} (\pi r_s/R)} {{\rm cos}(\pi r_s/R) +{\rm
cosh}(\sqrt{\frac{2\pi^3}{27}}\frac{|x^3|}{aR\kappa_0})}\Big{)}.
\end{eqnarray} 
After $r_s \rightarrow 0$ and multiplication 
by $4\pi R^2$ (to obtain the total emerging flux density at the surface) we
 obtain the equation given in equation~(\ref{eq:theoryspec})
(setting $R\kappa_0=\tau_0$):

\begin{equation}
J(x)=\frac{\sqrt{\pi}}{\sqrt{24}a\tau_0}\Bigg{(}\frac{x^2}{1+{\rm
cosh}\Big{[}\sqrt{\frac{2\pi^3}{27}}\frac{|x^3|}{a\tau_0}\Big{]}}\Bigg{)}
\hs\hs {\rm Solution \hs for\hs {\it sphere,}}
\end{equation} 
which can be compared to the previously known analytic solution for a
slab \citep{Harrington73,Neufeld90}:
\begin{equation}
J(x)=\frac{\sqrt{6}}{24\sqrt{\pi}a\tau_0}\Bigg{(}\frac{x^2}{1+{\rm
cosh}\Big{[}\sqrt{\frac{\pi^3}{54}}\frac{|x^3|}{a\tau_0}\Big{]}}\Bigg{)}
\hs \hs{\rm Solution \hs for \hs {\it slab.}}
\label{eq:theoryspecapp}
\end{equation}

\section{Deviations from Partial Coherence}
\label{sec:perturbations}

In \S~\ref{basics} we mentioned that for all applications discussed in
this paper, the scattering is partially coherent. The main reason for
this is that the Einstein $A$ coefficient is very high, which
translates to a very short spontaneous decay time, $t_{\rm spon} =
A_{21}^{-1}=10^{-8}$ s. Because the hydrogen atoms spend such a short
time in the $n=2$ state, the chance the atom gets perturbed, while in
this excited state, is tiny. Below this is quantified:

\subsection{Collisions with Electrons}
\label{sec:collisions}

Collisions between the excited hydrogen atom and electrons, can either
cause the energy of the re-emitted Ly$\alpha$ photon to differ from
that of the incoming photon, or result in the loss of the Ly$\alpha$
photon. The first process occurs when a collision puts the atom in any
state from which it radiatively decays back into the $2p$ state. The
last process occurs when the collision puts the atom in a state from
which the $2p$ state is not accessible via a radiative transition. An
example of such a state is the $2s$ level. \citet{Osterbrock62} gives
the collision rate from the 2p to 2s state, at $T=10^4$ \kel, $C_{\rm
2p2s}$ $\sim 2 \times 10^{-3}n_e$ s$^{-1}$, where $n_e$ is the number
density of electrons. This collision rate dominates the collision
rates into all other states by several orders of magnitude
\citep[e.g][]{Giovanardi87,Chang91}. Therefore, only the collision
rate from $2p$ to $2s$ is relevant for our purposes. Given that the
maximum densities in our models reach $n_e \sim 10^2$ cm$^{-3}$, the
rate at which excited hydrogen atoms get perturbed is at most
$10^{-1}$ s$^{-1}$, which is nine orders of magnitude below the
spontaneous decay rate, and therefore negligible.

\subsection{Absorption of Additional Photons}
\label{sec:photons}

The Ly$\alpha$ scattering rate for a given hydrogen atom is,

\begin{equation}
P_{\alpha}=4 \pi \int d \nu \frac{J_{\alpha}(\nu) \sigma(\nu)}{h \nu}
\hs {\rm s}^{-1} \approx \frac{4 \pi J_{\alpha}\sigma_0}{h \nu_0}\nu_0
\frac{v_{\rm th}}{c}\sim 5 \times 10^{-12}J_{\alpha,21}\hs {\rm
s}^{-1},
\end{equation} 
where we assumed that $J(x)=J_{\alpha}$ is constant and wrote
$J_{\alpha}=J_{\alpha,21}$ $\times 10^{-21}$ ergs s$^{-1}$ cm$^{-2}$
sr$^{-1}$ Hz$^{-1}$. $J_{\alpha,21}$ is highest in the central regions
in cases in which all Ly$\alpha$ is inserted at $r=0$. An estimate of
$J_{\alpha,21}$ can be obtained from:

\begin{equation}
J_{\alpha,21}=\frac{1}{4 \pi \Delta \nu}\frac{L_{\rm Ly \alpha}}{4 \pi
  r^2}= 3.3 \times 10^7 \Big{(} \frac{L_{\rm Ly \alpha}}{10^{43} \hs
  {\rm ergs} \hs {\rm s}^{-1}}\Big{)} \Big{(} \frac{1 \hs {\rm
    kpc}}{r}\Big{)}^2 \Big{(} \frac{10^{-4}}{ \Delta \nu
  /\nu_0}\Big{)},
\label{eq:scatrate}
\end{equation} 
where $\Delta \nu$ is the line width of the \lya line, which we set
conservatively to $10^{-4} \nu_0$. Our lines are typically much
broader, which would reduce $J_\alpha$. This implies a scattering rate
of $\sim 10^{-4}$ s$^{-1}$ at $r=1$ kpc.  Since the Einstein
B--coefficient for stimulated emission equals that of absorption, the
rate at which a hydrogen atom in the excited state absorbs \lya
photons equals that of a hydrogen atom in the ground state. The rate
at which stimulated emission of Ly$\alpha$ occurs per hydrogen atom is
therefore $\gtrsim 12$ orders of magnitude lower than $A_{21}$ for $r
> 1$ kpc, and can be safely ignored. 

In the presence of a continuum source , the photo-ionization rate per
hydrogen atom from the $2p$ state by any photon with $E>3.4$ eV,
emitted by a central source with a powerlaw spectrum $J(\nu) \propto
\nu^{-\beta}$ (for $h \nu > 3.4$ eV) is

\begin{equation}
P_{\rm ion}=4 \pi \int_{3.4 \hs {\rm
eV}}^{\infty}d\nu\frac{J(\nu)\sigma_{H_\alpha}(\nu)}{h\nu} \hs {\rm
s}^{-1}\approx 8 \times 10^{-8} \Big{(} \frac{L_{\rm ion}}{10^{45}\hs
{\rm ergs}\hs {\rm s}^{-1}}\Big{)}\Big{(}
\frac{\beta+1}{\beta+2.75}\Big{)}\Big{(} \frac{10 \hs {\rm
kpc}}{r}\Big{)}^2 \hs {\rm s}^{-1},
\end{equation} 
where $\sigma_{H_\alpha}(\nu)$ is the photo-ionization cross section
from the $n=2$ state at frequency $\nu$, which is $\sim 1.4 \times
10^{-17}(\nu/\nu_{\rm H \alpha})^{-2.75}$, where $h\nu_{\rm H
\alpha}=3.4$ eV \citep[e.g.][p.108]{Allen00}. The total luminosity of
the source is given by $L_{\rm tot}$. Photo-ionization out of the
$n=2$ state by continuum photons emitted by the central source is
clearly negligible. 

The final mechanism by which the atom may be perturbed in its excited
state is by absorbing other line photons. Since the Einstein $A$
coefficients are weaker for all other transitions in hydrogen, the
photo-excitation cross section to other levels is less than the
Ly$\alpha$ absorption cross section. For the same number density of
these other line photons, the photoexcitation rate is less than the
Ly$\alpha$ scattering rate calculated above, which was
negligible. Therefore, also this process can be safely ignored, which
completes our justification for only considering partially coherent
scattering (as compared to incoherent scattering).

\end{appendix}
\end{document}